\documentclass[sigconf, screen, nonacm]{acmart}

\setcopyright{none}

\usepackage{multirow}
\usepackage{tabularx}
\usepackage{enumitem}
\usepackage{xspace}
\usepackage{pifont}   
\usepackage{colortbl}
\usepackage{tikz}
\usetikzlibrary{positioning,arrows.meta,fit,calc,patterns}
\usepackage{fontawesome5}   

\setlist{leftmargin=*}

\newcommand{\sys}{\texttt{Mayfly}\xspace}          

\newcommand{\need}[1]{\textbf{D#1}}       
\definecolor{oursblue}{HTML}{2C4E86}      
\definecolor{oursbar}{HTML}{7799D2}
\definecolor{oursrow}{HTML}{E4EBF6}       
\definecolor{insightbg}{HTML}{E4E4E4}     
\definecolor{insightbar}{HTML}{4D4D4D}    
\definecolor{readybg}{HTML}{FCE2CB}       
\definecolor{parkedbg}{HTML}{DCE4ED}      

\AtBeginDocument{\ifdefined\pdfstringdefDisableCommands\pdfstringdefDisableCommands{}\fi}

\tikzset{boxstyle/.style={fill=insightbg, rounded corners=3pt, inner xsep=7pt, inner ysep=4pt,
  text width=\dimexpr\linewidth-14pt\relax, align=justify}}
\newcommand{\obsbox}[2]{
  \par\smallskip\noindent
  \begin{tikzpicture}
    \node[boxstyle] (tk) {\textbf{\color{insightbar}Takeaway~#1.}\ #2};
    \begin{scope}\clip[rounded corners=3pt] (tk.south west) rectangle (tk.north east);
      \fill[insightbar] (tk.south west) rectangle ([xshift=2.5pt]tk.north west);\end{scope}
  \end{tikzpicture}%
  \par\smallskip}

\newcommand{\x}{$\times$\xspace}

\begin{document}

\title{\texttt{Mayfly}: Lifecycle-Aware High-Bandwidth Flash for Recursive Self-Improvement}   

\author{Chunlin Tian}
\affiliation{\institution{National University of Singapore}\country{Singapore}}
\author{Mingjia Fan}
\affiliation{\institution{National University of Singapore}\country{Singapore}}
\author{Feng Yu}
\affiliation{\institution{National University of Singapore}\country{Singapore}}
\author{Qingyun Zou}
\affiliation{\institution{National University of Singapore}\country{Singapore}}
\author{Wenju Zhao}
\affiliation{\institution{Huazhong University of Science and Technology}\country{China}}
\author{Weng-Fai Wong}
\affiliation{\institution{National University of Singapore}\country{Singapore}}
\author{Bingsheng He}
\authornote{Corresponding author.}
\affiliation{\institution{National University of Singapore}\country{Singapore}}
\renewcommand{\shortauthors}{Tian et al.}

\begin{abstract}
Recursive self-improvement (RSI) lets an LLM improve its agent system by proposing and evaluating candidate variants across tasks. A generation's sessions grow their KV caches together and release them as they finish, creating a burst followed by a decline. At high concurrency, even the KV of sessions ready to run exceeds the GPU's HBM. Offloading that KV to host DRAM or SSDs adds capacity but leaves ready sessions waiting for HBM space before execution. High-Bandwidth Flash (HBF) can serve the overflow in place, but excessive reads lengthen decode steps and repeated writes consume flash endurance. Our key insight is that where ready KV is read should follow the burst and its decline: from HBF in place while the burst overflows HBM, as much as each step's bandwidth allows, and from HBM again as the decline frees it. A switch between ready and parked decides only whether a block is read, so its flash copy stays valid, and a session's end, which the orchestrator announces, decides when its private KV dies. We present \sys, lifecycle-guided HBF management for RSI. \sys bounds per-step HBF reads during overflow and returns KV to HBM as sessions finish and free space. It retains a valid backing copy of each block evicted from HBM to avoid rewrites across state transitions, and forwards session-end signals to the controller to avoid copying dead KV during garbage collection. Experimental results show that, at 128 sessions per generation on four added HBF sites, \sys finishes 34\% sooner than parking the same KV on HBF and 23\% sooner than reading it from HBF whenever HBM is full; its gain over parking grows with concurrency, to 55\% at 256 sessions on eight sites. Keeping one flash copy per block and trimming at announced session ends avoid write amplification.
\end{abstract}

\maketitle

\section{Introduction}
\label{sec:intro}

Recursive self-improvement (RSI) loops let an LLM improve the agent system it runs in. We study harness RSI, which keeps
the model fixed, rewrites its harness, the prompts, tools and control flow, and scores each candidate by
running it~\cite{dgm,hgm,adas,godelagent}: a generation evaluates a batch of candidates on a set of tasks, and each
evaluation is an agent session that alternates LLM inference with tool calls~\cite{swebench,swerebench}. Selection
waits for every session, so the time to finish a generation sets the loop's pace. The serving
engine keeps each session's KV cache across its calls, and the sessions compete for the HBM left after model
weights~\cite{vllm,sglang}. In the loops we study, the orchestrator starts a generation's sessions together and does not
replace those that finish, so their KV bursts and then declines (Fig.~\ref{fig:burst}a). Within the burst, a session is
\emph{parked} while it waits for a tool and \emph{ready} while it has an LLM call running or
queued, wherever its KV is placed. As contention for the GPU and its HBM lengthens LLM calls, more of the KV is ready: in a trace-driven
simulation of 32 RSI~\cite{dgm} sessions per generation on one H200 that holds KV in HBM only, ready KV alone exceeds HBM 86\% of the run time,
against 55\% for all live KV when no call waits for HBM (Fig.~\ref{fig:burst}a), as queued calls keep their sessions ready (\S\ref{sec:char-load}).

\begin{figure}[t]
  \centering
\begin{tikzpicture}[x=1cm, y=1cm, line join=round,
    every node/.style={font=\fontfamily{ptm}\fontsize{7}{8.2}\selectfont, inner sep=0pt, outer sep=0pt},
    mf blk/.style={draw=mfline, line width=0.3pt},
    note/.style={text=black, font=\fontfamily{ptm}\fontsize{7}{8}\selectfont},   
  ]
  \definecolor{mfline}{HTML}{404040}
  \definecolor{mfready}{HTML}{FFC6A9}
  \definecolor{mfpark}{HTML}{C2D1EB}
  \definecolor{mfkv}{HTML}{D9D9D9}
  \definecolor{mfgrey}{HTML}{6E6E6E}
  \definecolor{mfrule}{HTML}{C8C8C8}
  \def\sess#1#2#3{%
    \fill[mfpark] (#1,#2) rectangle (#1+#3,#2+0.12);
    \foreach \a/\b in {0/0.17, 0.30/0.46, 0.62/0.76, 0.92/1.08, 1.24/1.38, 1.54/1.70, 1.86/2.00} {
      \pgfmathsetmacro{\aa}{min(\a,#3)}\pgfmathsetmacro{\bb}{min(\b,#3)}
      \fill[mfready] (#1+\aa,#2) rectangle (#1+\bb,#2+0.12);}
    \draw[mf blk] (#1,#2) rectangle (#1+#3,#2+0.12);}%
  \def\seg#1#2#3#4{\draw[mf blk, fill=#4] (#1,#3) rectangle (#2,#3+0.14);}%
  \def\hd#1#2{\node[font=\fontfamily{ptm}\fontsize{7}{8.4}\selectfont\bfseries] at (#1,2.44) {#2};}%

  \hd{2.80}{who starts sessions}
  \hd{5.25}{next turn}
  \hd{7.40}{KV dies}
  \draw[draw=mfrule, line width=0.5pt] (0,2.27) -- (8.45,2.27);
  \draw[draw=mfrule, line width=0.5pt] (0,1.20) -- (8.45,1.20);
  \node[anchor=west, align=left, font=\fontfamily{ptm}\fontsize{7}{8.4}\selectfont\bfseries] at (0,1.735) {Agent\\serving};
  \node[anchor=west, align=left, font=\fontfamily{ptm}\fontsize{7}{8.4}\selectfont\bfseries] at (0,0.62) {Harness\\RSI};

  \node[text=mfgrey, font=\fontsize{10}{11}\selectfont] at (1.50,1.77) {\faIcon{users}};
  \sess{1.85}{1.98}{0.95}
  \sess{2.17}{1.80}{0.90}
  \sess{2.53}{1.62}{1.00}
  \sess{2.90}{1.44}{0.95}
  \node[text=mfgrey, font=\fontsize{10}{11}\selectfont] at (1.50,0.65) {\faIcon{sitemap}};
  \sess{1.85}{0.86}{0.95}
  \sess{1.85}{0.68}{1.25}
  \sess{1.85}{0.50}{0.75}
  \sess{1.85}{0.32}{2.15}
  \draw[draw=mfline, line width=0.5pt, dash pattern=on 1.4pt off 1pt] (1.85,0.24) -- (1.85,1.04);

  \seg{4.15}{4.65}{1.75}{mfready}
  \draw[draw=mfline, line width=0.5pt, dash pattern=on 1.4pt off 1pt] (4.65,1.82) -- (5.85,1.82);
  \node[font=\fontfamily{ptm}\fontsize{7}{8}\selectfont\bfseries] at (5.25,1.97) {?};
  \draw[draw=mfline, line width=0.3pt, dash pattern=on 1.4pt off 1pt, fill=white] (5.85,1.75) rectangle (6.20,1.89);
  \node[note] at (5.18,1.53) {user may not return};
  \seg{4.15}{4.65}{0.62}{mfready}
  \seg{4.65}{5.35}{0.62}{mfpark}
  \seg{5.35}{5.85}{0.62}{mfready}
  \node[note, anchor=south] at (5.00,0.80) {tool call};
  \node[note] at (5.18,0.42) {tool returns, resumes};

  \fill[mfkv] (6.55,1.75) rectangle (7.25,1.89);
  \fill[pattern=north east lines, pattern color=mfgrey] (7.25,1.75) rectangle (8.20,1.89);
  \draw[mf blk] (6.55,1.75) rectangle (8.20,1.89);
  \draw[draw=mfline, line width=0.9pt] (8.20,1.68) -- (8.20,1.96);
  \node[note, anchor=south] at (8.20,1.99) {TTL};
  \draw[draw=mfline, line width=0.5pt] (7.25,1.70) -- (7.25,1.94);
  \node[note, anchor=south] at (7.25,1.99) {last turn};
  \node[note] at (7.38,1.53) {maybe dead, still kept};
  \fill[mfkv] (6.55,0.62) rectangle (7.25,0.76);
  \draw[mf blk] (6.55,0.62) rectangle (7.25,0.76);
  \draw[draw=mfline, line width=0.9pt] (7.25,0.55) -- (7.25,0.83);
  \node[note, anchor=south] at (7.25,0.86) {session ends};
  \node[note] at (7.38,0.42) {dead at once};
\end{tikzpicture}%
  \caption{KV lifetimes in RSI and in agent serving. Bars are sessions (peach: LLM call, blue: tool call); grey: KV lifetime.}
  \label{fig:life}
\end{figure}
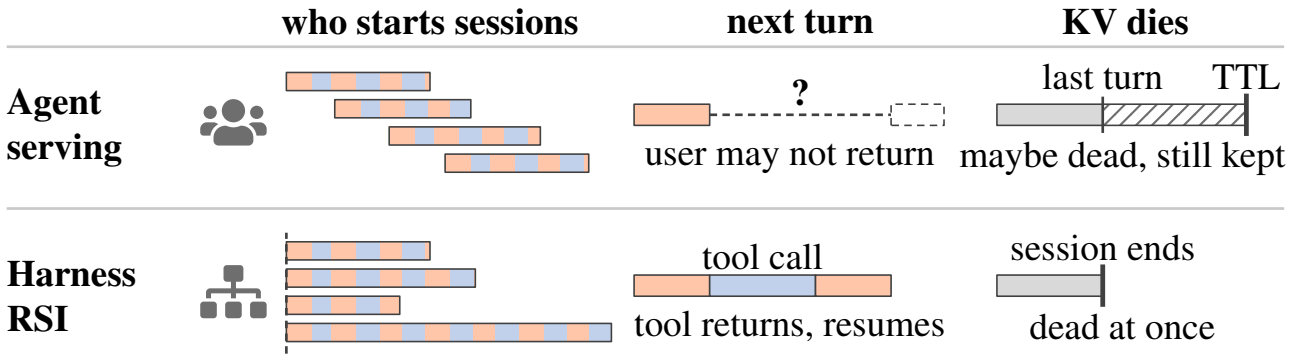

\textbf{State of the Art.} Agent-serving systems restore such KV from host memory or drives before its session
runs~\cite{cachedattention,mooncake,lmcache,infercept,kvflow,continuum}, adding capacity but not room in HBM.
High-bandwidth flash (HBF) puts NAND stacks on the GPU package~\cite{ocphbf,patterson_hbf}, so KV beyond HBM can take
part in a step where it lies, at limited sustained read bandwidth and write endurance. Recent work uses it for agentic
and LLM serving, holding paused sessions' KV and restoring it~\cite{baek_hbf} or letting decode read KV in
place~\cite{flashaccel,hbflex,yu_hbf,seta,splash}. Whether flash suits KV is disputed: a server that is not told when a
session ends cannot tell when its KV dies, and in production serving traces each byte written to a KV pool is read
back only 0.20--0.88 times~\cite{hbfsucks}. Lifetime-aware placement goes further: it picks each context segment's tier at allocation and frees blocks at context boundaries the harness reports~\cite{lachesis}. A
placement made when KV is written, however, does not decide how the KV is served at later steps, nor use the HBM that
finishing sessions free: even when a block's end is known, whether to read it in place or move it back depends on the
current ready set, the free HBM and the step's bandwidth.\textit{ Thus, how should HBF serve RSI's ready KV
as it grows beyond HBM and then shrinks, so that ready sessions stop waiting for HBM space while flash reads do not
lower step throughput and flash is written no more than the KV itself requires?}

\textbf{Challenges.} Each phase of the burst works against one of these goals. As the burst rises, a generation's
sessions overflow together: if every ready session reads its overflow from HBF at every step, the reads exceed a
stack's sustained bandwidth and lengthen every step. As it declines, finishing sessions free HBM, and KV still read
from flash once HBM has room again costs time and energy. Throughout, a session switches between ready and parked
dozens of times, mostly for under a second, so a tier that moves KV at each switch rewrites flash every time, and one
that keeps KV for a fixed time makes garbage collection copy KV whose session has ended.

 \textbf{Our Contribution.} \textbf{To our knowledge, this is the first work to manage HBF for recursive self-improvement, whose KV bursts past HBM and then declines.} RSI exposes the states and events that these decisions need. Our key insight is that where ready KV is read should follow the burst and its decline: from HBF in place while the burst overflows HBM, as much as each step's bandwidth allows, and from HBM again as the decline frees it. A switch between ready and parked decides only whether a block is read, so its flash copy stays valid, and a session's end, which the orchestrator announces, decides when its private KV dies. Unlike interactive agent serving, where KV waits on a timer for a user who
may not return, the orchestrator ends every session itself (Fig.~\ref{fig:life}): an RSI session ends when its agent
replies without a tool call and is never resumed, so its private KV is dead at once, while a shared task prefix lives until its last user ends. We present \sys, which turns these
signals into lifecycle-guided HBF management. A per-step gate lets a request read KV in place only if this does not
lower the step's estimated token throughput, and each admitted request attends over its whole context; as sessions
finish and free HBM, \sys moves KV back so that later steps stop reading it from flash. \sys keeps a valid HBF copy of
each block evicted from HBM, so later evictions write nothing, and forwards session ends to the HBF controller, so that
garbage collection copies none of a finished session's private KV. It spans the serving runtime, a two-tier attention
kernel and a KV-aware controller on the HBF base die.
We evaluate \sys on real agent traces and recorded RSI runs in a discrete-event simulator whose GPU model is calibrated
against an H200 and whose HBF path is modelled. With the same four added HBF sites, at 128 sessions per generation, \sys finishes
34\% sooner than parking the same KV on HBF, with a 1.9\x lower median time to first token, and 23\% sooner than reading it from HBF whenever HBM is full; at 256 sessions, on eight sites, it finishes 55\% sooner than parking, with a 2.6\x lower median time to first token. Even parked on HBF, RSI's KV is read back 23--28 times per byte written, against 0.20--0.88 in serving KV pools.
We make the following contributions:
\begin{itemize}[itemsep=1pt,topsep=3pt]
\item A characterization: under load, ready sessions hold most of RSI's KV, and their KV alone exceeds HBM
      most of the time (\S\ref{sec:char}).
\item \sys, which uses session states, free HBM and session ends to decide which ready KV each step reads in place
      or returns to HBM, and to write each block to flash once from the host (\S\ref{sec:design}).
\item A wear analysis showing that one copy per block and trims at session ends avoid write amplification: a stack
      lasts 1.7--2.8\x as long as with a 30-min expiry or release when the tier is full, and 9--37\x as long as when
      each return to HBM drops its copy (\S\ref{sec:eval}).
\end{itemize}

\section{Background and Related Work}
\label{sec:bg}
\label{sec:related}

\subsection{Recursive Self-Improvement}
\label{sec:bg-rsi}

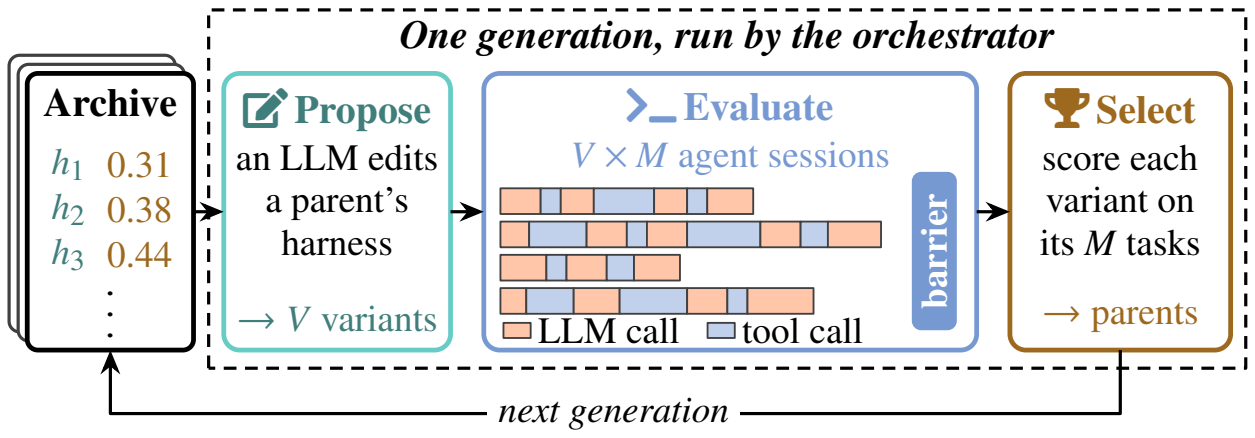
\begin{figure}[t]
  \centering
\begin{tikzpicture}[x=1cm, y=1cm, line join=round,
    every node/.style={font=\fontfamily{ptm}\fontsize{7.5}{9}\selectfont, inner sep=0pt, outer sep=0pt},
    mf phase/.style={draw=#1, line width=0.9pt, rounded corners=3pt, fill=white},
    mf blk/.style={draw=mfline, line width=0.35pt},
    mf arr/.style={line width=0.6pt, -{Stealth[length=1.6mm,width=1.3mm]}},
    mf txt/.style={align=center, font=\fontfamily{ptm}\fontsize{7}{8}\selectfont},   
  ]
  \definecolor{mfline}{HTML}{404040}
  \definecolor{mfready}{HTML}{FFC6A9}
  \definecolor{mfpark}{HTML}{C2D1EB}
  \definecolor{mfgreen}{HTML}{68CCC6}
  \definecolor{mfgreendk}{HTML}{407E7B}
  \definecolor{mfblue}{HTML}{7799D2}
  \definecolor{mfgold}{HTML}{9D691F}
  \def\mfS{\fontsize{7}{8}\selectfont}%
  \def\mfL{L}%
  \def\turns#1#2{\foreach \xa/\xb/\typ in {#2} {
      \ifx\typ\mfL\def\clr{mfready}\else\def\clr{mfpark}\fi
      \draw[mf blk, fill=\clr] (\xa,#1) rectangle (\xb,#1+0.17);}}%

  \foreach \d in {0.12,0.06}
    \draw[line width=0.6pt, draw=mfline, rounded corners=2pt, fill=white] (0.12-\d,0.72+\d) rectangle (1.22-\d,2.55+\d);
  \draw[line width=0.9pt, draw=black, rounded corners=2pt, fill=white] (0.12,0.72) rectangle (1.22,2.55);
  \node[anchor=north] at (0.67,2.45) {\textbf{Archive}};
  \foreach \y/\h/\s in {1.95/1/0.31, 1.65/2/0.38, 1.35/3/0.44} {
    \node[anchor=west] at (0.28,\y) {\mfS\color{mfgreendk}\textit{h}\textsubscript{\h}};
    \node[anchor=east] at (1.08,\y) {\mfS\color{mfgold}\s};
  }
  \node at (0.67,1.06) {\mfS$\vdots$};

  \draw[line width=0.6pt, dash pattern=on 2.4pt off 1.6pt] (1.32,0.6) rectangle (8.18,2.97);
  \node[anchor=north] at (4.75,2.9) {\bfseries\itshape One generation, run by the orchestrator};

  \draw[mf phase=mfgreen] (1.42,0.72) rectangle (2.92,2.55);
  \node[anchor=north, text=mfgreendk] at (2.17,2.45) {\faIcon{edit}\,\textbf{Propose}};
  \node[mf txt, anchor=north] at (2.17,2.08) {an LLM edits\\a parent's\\harness};
  \node[mf txt, anchor=south, text=mfgreendk] at (2.17,0.84) {$\rightarrow$ \textit{V} variants};

  \draw[mf phase=mfblue] (3.14,0.72) rectangle (6.4,2.55);
  \node[anchor=north, text=mfblue] at (4.77,2.45) {\faIcon{terminal}\,\textbf{Evaluate}};
  \node[mf txt, anchor=north, text=mfblue] at (4.77,2.1) {\textit{V}\,$\times$\,\textit{M} agent sessions};
  \begin{scope}[xshift=3.25cm, xscale=0.95]
    \turns{1.62}{0/.28/L, .28/.42/T, .42/.65/L, .65/1.07/T, 1.07/1.30/L, 1.30/1.44/T, 1.44/1.76/L}
    \turns{1.40}{0/.20/L, .20/.60/T, .60/.88/L, .88/1.02/T, 1.02/1.30/L, 1.30/1.81/T, 1.81/2.09/L, 2.09/2.28/T, 2.28/2.65/L}
    \turns{1.18}{0/.32/L, .32/.46/T, .46/.74/L, .74/.93/T, .93/1.25/L}
    \turns{0.96}{0/.18/L, .18/.51/T, .51/.83/L, .83/1.30/T, 1.30/1.58/L, 1.58/1.72/T, 1.72/2.18/L}
  \end{scope}
  \draw[line width=0.5pt, rounded corners=1.5pt, fill=mfblue, draw=mfblue] (5.98,0.86) rectangle (6.26,1.89);
  \node[rotate=90, text=white] at (6.12,1.375) {\mfS\bfseries barrier};
  \draw[mf blk, fill=mfready] (3.27,0.8) rectangle (3.45,0.9);
  \node[anchor=west] at (3.50,0.85) {\fontsize{7}{8}\selectfont LLM call};
  \draw[mf blk, fill=mfpark] (4.62,0.8) rectangle (4.80,0.9);
  \node[anchor=west] at (4.85,0.85) {\fontsize{7}{8}\selectfont tool call};

  \draw[mf phase=mfgold] (6.62,0.72) rectangle (8.08,2.55);
  \node[anchor=north, text=mfgold] at (7.35,2.45) {\faIcon{trophy}\,\textbf{Select}};
  \node[mf txt, anchor=north] at (7.35,2.08) {score each\\variant on\\its \textit{M} tasks};
  \node[mf txt, anchor=south, text=mfgold] at (7.35,0.84) {$\rightarrow$ parents};

  \draw[mf arr] (1.22,1.635) -- (1.42,1.635);
  \draw[mf arr] (2.92,1.635) -- (3.14,1.635);
  \draw[mf arr] (6.4,1.635) -- (6.62,1.635);
  \draw[mf arr] (7.35,0.72) -- (7.35,0.3) -- (0.67,0.3) -- (0.67,0.72);
  \node[anchor=center, fill=white, inner xsep=2pt, inner ysep=1pt] at (4.0,0.3) {\mfS\itshape next generation};
\end{tikzpicture}%
  \caption{One generation of a harness RSI loop. Selection waits for all of its sessions to end.}
  \label{fig:rsiloop}
\end{figure}

In recursive self-improvement (RSI), an AI system changes itself, evaluates the change, and keeps what
improves~\cite{good1965,godelmachine}; with LLMs, a model writes the change and a benchmark judges it~\cite{stop,adas}.
A loop may change the model's weights~\cite{rzero}, a task's prompts or solution
programs~\cite{promptbreeder,funsearch,alphaevolve,openevolve}, or the \emph{harness} around a fixed model: its
prompts, tools, memory and control flow~\cite{dgm,godelagent,hgm,metaharness,ahe,rrsi,harness2,sift}. We study harness
loops and call them RSI below. With fixed weights, their GPU serves inference, and the memory they add is the KV that
concurrent sessions keep across calls.
Each generation runs three steps (Fig.~\ref{fig:rsiloop}): it \emph{proposes} $V$ candidate harnesses,
\emph{evaluates} each on $M$ tasks, and \emph{selects} the candidates that seed the next generation. Each evaluation
runs one \emph{session}, in which an agent built on one candidate solves one task, such as a GitHub issue, over many
turns. Each turn issues a \emph{request}, one LLM call, whose output usually invokes a tool in a sandbox, and the
tool's result enters the next request. A session is \emph{ready} while it has a request running or queued and
\emph{parked} while it waits for a tool, and so is its KV, wherever it is placed. In the loops we study, a generation's sessions
start together and selection waits for all to end.
The serving engine keeps a session's KV between requests, so that each request prefills only the tokens added since
the last one~\cite{vllm,sglang}. Three properties of this KV matter below. A request's end does not invalidate it, since the session's next request
reads its history again; a switch between ready and parked changes only whether it is read; and the orchestrator's
session end makes a session's private KV reclaimable. A stored copy stays reusable only while the context it encodes is
unchanged: a harness that appends keeps it valid, whereas trimming or compressing the context invalidates the KV from
the first changed token on.

\subsection{High-Bandwidth Flash}
\label{sec:bg-hbf}

High-bandwidth flash (HBF) stacks NAND dies on a base die behind a wide interface, so that a stack can sit beside
HBM on the GPU package~\cite{ocphbf,patterson_hbf}. We call a stack's position on the package a
\emph{site}: an H200 has six HBM sites, and each HBF stack takes a site of its own. Capacity comes from NAND density:
the OCP specification targets 550\,GB in 16 dies, against 48--64\,GB for an HBM4 stack~\cite{ocphbf}. Bandwidth
comes from array parallelism: a page read takes microseconds, so a stack reaches high bandwidth only by keeping many
planes busy at once~\cite{hbflex,splash}, and the rate it sustains under power and thermal limits lies below its
interface's peak~\cite{hbfsucks}.
HBF keeps NAND's management constraints. A page is programmed once and erased only with its whole block, so garbage
collection (GC) copies a block's still-valid pages before erasing it; the write amplification factor (WAF) counts these
copies. A cell endures a limited number of program/erase (P/E) cycles. A controller not told that data is dead copies it
in every collection that reaches it.

%
%

\subsection{Related Work}
\label{sec:bg-offload}

\begin{figure*}[t]
  \centering
  \includegraphics[width=\textwidth]{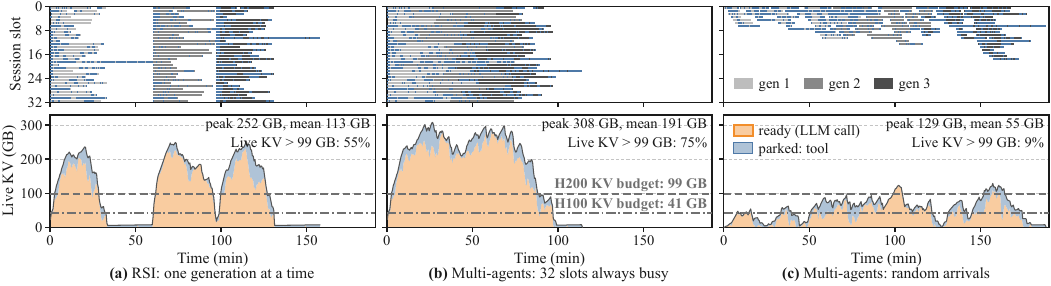}
  \caption{Sessions of three generations (4 variants $\times$ 8 tasks) under three arrival patterns. Top: sessions by slot,
    tool calls in blue. Bottom: live KV, with unlimited memory, against the H200 and H100 KV budgets.}
  \label{fig:burst}
\end{figure*}

\noindent \textbf{Conventional KV offloading.} Agent serving keeps a session's KV across calls because prefilling a
long history again at every turn costs more GPU time than storing it~\cite{cachedattention,infercept}. When HBM cannot
hold this KV, systems move it to host DRAM, local SSDs or remote memory pools. They restore
histories into HBM at a turn's start~\cite{cachedattention,mooncake,lmcache,impress,tutti,pykvcache}, keep, swap or
prefetch KV along the agents' schedule~\cite{infercept,kvflow,continuum}, and limit what they write to
SSDs~\cite{gandalf,saria}. In these restore-based schemes, the far tier adds capacity, but a session runs only once
its KV is back in HBM, so HBM's space bounds the ready sessions. A second line computes attention where the KV lies,
on host CPUs or inside the drive~\cite{neo,instinfer,hilos}; it skips the restore and is bound by the host's or
drive's bandwidth.

\noindent \textbf{HBF-backed LLM and agent serving.} HBF brings flash capacity onto the package, and recent work
uses it for KV in two ways. Hot--cold tiering parks paused agent sessions' KV on HBF while hot KV stays in HBM, and
restores a session's KV when it resumes~\cite{baek_hbf}. Execution-tier designs let decode read KV from HBF in place,
all of it~\cite{flashaccel,hbflex,yu_hbf,seta} or a budgeted subset of older pages~\cite{splash}, so that long
contexts and large batches run beyond HBM's capacity. Two keep a block's flash copy valid while it also sits in
HBM~\cite{baek_hbf,flashaccel}. HBF has also been studied for general LLM inference~\cite{son_hbf} and for
generative recommendation with a write-aware KV policy~\cite{peng_hbf}; MemExplorer's design search fits it to
long-context KV overflow without modeling write endurance~\cite{memexplorer}.

\noindent \textbf{Is HBF suited to KV?} Whether KV belongs on flash turns on three questions. First, whether faster far access removes
the bottleneck: if ready sessions wait for HBM space, a faster restore leaves the wait. Second, whether
KV written to flash is read enough to pay for its writes: KV grows with every generated token~\cite{patterson_hbf}, so
a flash tier that holds it receives a stream of appends. Third, whether the bandwidth holds under power, thermal and endurance limits, where
on-package LPDDR may suit KV better~\cite{petrucci_hbf}. Li et al.~\cite{hbfsucks} replace the backing tier of
SSD-style KV pooling with HBF and find that the pool absorbs more bytes of writes than it returns to HBM and that thermal
limits cap the stack's bandwidth; they question HBF as a drop-in replacement for that tier, not other uses.
Lifetime-aware placement~\cite{lachesis} cuts HBF writes by placing each context segment at allocation, a decision
that covers neither later reads nor migration as HBM frees. HBF's effect thus depends on which KV reaches flash, how it is read later,
and whether the device learns when it dies.

\section{Characterization and Motivation}
\label{sec:char}

\subsection{Ready KV Outgrows HBM}
\label{sec:char-agents}
\label{sec:char-time}
\label{sec:char-load}

\begin{figure}[t]
  \centering
  \includegraphics[width=\columnwidth]{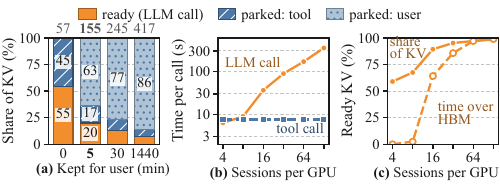}
  \caption{(a)~TraceLab agents' KV on the busiest day by state and user wait; mean GB on top. (b,~c)~The sessions
    of Fig.~\ref{fig:burst}a on one H200 with HBM only, under rising load.}
  \label{fig:load}
\end{figure}

We study RSI's KV on one H200 in two ways: we run sessions with vLLM, and we replay recorded sessions in a
discrete-event simulator of vLLM whose step times are fitted on H200 replays, to vary the load, arrivals and memory.

\noindent \textbf{Burst and decline.} Interactive agent serving runs unrelated users' sessions, which start at any time; an RSI
orchestrator starts a generation's sessions together and runs every tool call, so the next turn comes as soon as a
tool returns. Figure~\ref{fig:burst} simulates the same 96 sessions of a real RSI run~\cite{dgm} on one H200 under three arrival patterns. As
an RSI loop (Fig.~\ref{fig:burst}a), a generation's 32 sessions grow their contexts together, and KV peaks at
236--252\,GB in the three generations. Finished sessions are not replaced, so KV then falls while the generation waits for its slowest; in
generations~1 and~3, the last session runs alone for 27 of 60\,min. The overflow thus changes within a generation: every
session that ends frees HBM, so KV that overflows HBM at the peak fits there again later. As multi-agents that keep all 32 slots busy
(Fig.~\ref{fig:burst}b), the same sessions keep KV near its 191\,GB mean, and arriving at random
(Fig.~\ref{fig:burst}c), they exceed the H200 KV budget 9\% of the time, against 55\% for RSI. Real loops show both
phases: the real DGM run peaks at 3.1\x its mean KV as each generation's evaluation starts, and our loaded loop, real
OpenHands~\cite{openhands} runs on SWE-rebench issues~\cite{swerebench} in DGM's structure, peaks at 239\,GB, 2.4\x the KV that HBM holds.

\noindent \textbf{Without users, parked KV waits only for tools.} In interactive agent serving, a session is also \emph{parked} while it waits for its user: in
TraceLab's 8,058 Claude Code and Codex sessions~\cite{tracelab}, users
reply after 130\,s at the median and after more than 5\,min in 25\% of the gaps. On the busiest day, with KV kept for
the 5\,min of hosted prompt caches, sessions hold 155\,GB on average, of which 20\% is ready, 17\% waits for tools
and 63\% for users (Fig.~\ref{fig:load}a), and their ready KV stays within an H200's HBM 98\% of the time, so agent
serving can park most of its KV in host memory or on drives~\cite{infercept,kvflow,continuum,lmcache}. Harness RSI
has no users: in the first bar, 55\% of its KV is ready and the remaining 45\% waits for tools to return.

\noindent \textbf{Load makes most of the KV ready.} An RSI operator raises the load, the sessions a loop runs at
once (slots) and the loops that share a GPU, to finish more sessions per hour. Figure~\ref{fig:load}b,~c simulates
Fig.~\ref{fig:burst}a's sessions with KV in HBM only, where vLLM recomputes what does not fit, one loop at 4 to 32 slots or two and four such loops started 900\,s apart; throughput peaks at 27 sessions per
hour at 8 slots, two per variant, and falls to 14--15 from 32 slots on. Tools keep their recorded 7.6\,s on average, while an LLM call stretches from 6.3\,s at 4 slots to 89\,s at 32 and 352\,s at four
loops, as each step serves more sessions and preempted ones wait for HBM (Fig.~\ref{fig:load}b). The ready share grows from 59\% to
95\% and 99\%, and ready KV alone exceeds HBM 86 and 99\% of the time (Fig.~\ref{fig:load}c), averaging 166 and
560\,GB against the H200's 99\,GB KV budget.

\obsbox{1}{RSI's KV bursts beyond HBM and declines as sessions end. Under load, its ready KV alone exceeds HBM, whereas
agents' kept KV mostly waits for users.}

\subsection{Restoring and Unbounded Reads Cost Time}
\label{sec:whyhbf}
\label{sec:char-capacity}
\label{sec:char-tier}
\label{sec:char-models}
\label{sec:align}

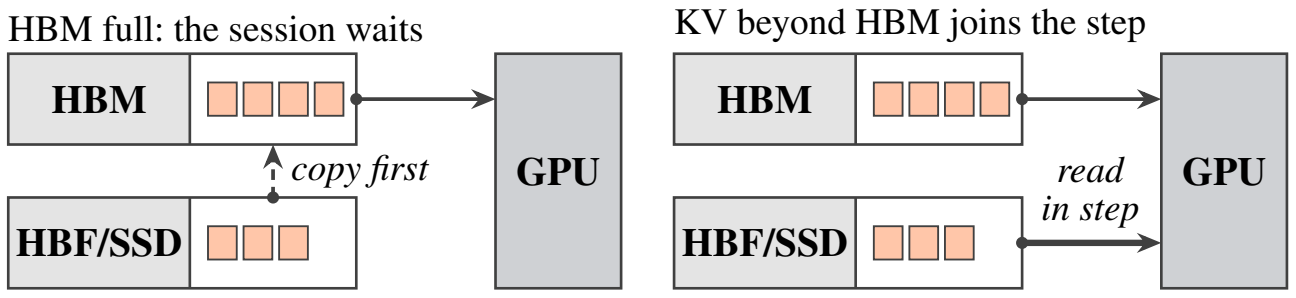
\begin{figure}[t]
  \centering
  \resizebox{\columnwidth}{!}{
\begin{tikzpicture}[x=1cm, y=1cm, line join=miter,
    every node/.style={font=\fontfamily{ptm}\fontsize{7.5}{9}\selectfont, inner sep=0pt, outer sep=0pt},
    sw box/.style={draw=mfline, line width=0.5pt},
    sw arr/.style={line width=0.7pt, draw=mfline, -{Stealth[length=1.6mm,width=1.3mm]}},
    sw note/.style={font=\fontfamily{ptm}\fontsize{7}{8.2}\selectfont},
  ]
  \definecolor{mfline}{HTML}{404040}
  \definecolor{mfready}{HTML}{FFC6A9}
  \definecolor{mfheld}{HTML}{E4E4E4}
  \definecolor{swgpu}{HTML}{CFD1D4}
  \def\tier#1#2#3#4{%
    \fill[white] (#1,#2) rectangle (#1+2.3,#2+0.6);
    \fill[mfheld] (#1,#2) rectangle (#1+1.2,#2+0.6);
    \draw[sw box] (#1,#2) rectangle (#1+2.3,#2+0.6);
    \draw[sw box] (#1+1.2,#2) -- (#1+1.2,#2+0.6);
    \node at (#1+0.6,#2+0.3) {\textbf{#3}};
    \foreach \i in {1,...,#4} {
      \fill[mfready] (#1+1.085+\i*0.235,#2+0.19) rectangle ++(0.19,0.22);
      \draw[draw=mfline, line width=0.35pt] (#1+1.085+\i*0.235,#2+0.19) rectangle ++(0.19,0.22);}}%
  \def\panel#1#2#3#4#5{%
    \tier{#1}{1.25}{HBM}{4}
    \tier{#1}{0.3}{HBF/SSD}{3}
    \fill[swgpu] (#1+3.22,0.3) rectangle (#1+4.05,1.85);
    \draw[sw box] (#1+3.22,0.3) rectangle (#1+4.05,1.85);
    \node at (#1+3.635,1.075) {\textbf{GPU}};
    \fill[mfline] (#1+2.3,1.55) circle (0.04);
    \draw[sw arr] (#1+2.3,1.55) -- (#1+3.22,1.55);
    \node[sw note, anchor=south west] at (#1,1.93) {#5};
    \node[anchor=north, align=center, font=\fontfamily{ptm}\fontsize{7}{8.4}\selectfont] at (#1+1.98,0.2) {\textbf{#2} #3: #4};}%

  \panel{0}{(a)}{Swap}{copy to HBM, then compute}{HBM full: the session waits}
  \fill[mfline] (1.75,0.9) circle (0.04);
  \draw[sw arr, dash pattern=on 2pt off 1.3pt] (1.75,0.9) -- (1.75,1.25);
  \node[sw note, anchor=west] at (1.87,1.075) {\itshape copy first};

  \panel{4.40}{(b)}{In place}{read the far tier in the step}{KV beyond HBM joins the step}
  \fill[mfline] (6.70,0.6) circle (0.04);
  \draw[sw arr, line width=1.2pt] (6.70,0.6) -- (7.62,0.6);
  \node[sw note, anchor=south, align=center] at (7.15,0.7) {\itshape read\\[-1pt]\itshape in step};
\end{tikzpicture}%
}
  \caption{KV beyond HBM reaches the GPU (a)~swapped into HBM before its session runs or (b)~read in place.}
  \label{fig:swap}
\end{figure}

We run a generation of 32 repository-level sessions with vLLM on one H200. It keeps up to 261\,GB of KV alive. With HBM alone,
vLLM recomputes 37.3M tokens of prefill and takes 122\,min and 4.1\,MJ, against 31\,min and 0.9\,MJ with 384\,GB of
host memory for offload. A larger HBM only moves the threshold: in simulation, a B300 holds a 32-session generation
at 0.95\x its KV capacity and 40 sessions need 1.3\x. KV beyond HBM reaches the GPU in two ways (Fig.~\ref{fig:swap}), and both cost time.
We compare them, and in \S\ref{sec:char-life} two write policies, in this simulator.

\begin{figure}[t]
  \centering
  \includegraphics[width=\columnwidth]{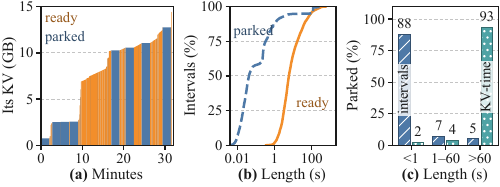}
  \caption{(a)~One session's KV. (b)~Ready and parked interval lengths. (c)~Parked intervals and KV-time by length.}
  \label{fig:switch}
\end{figure}

\noindent \textbf{Restoring makes ready sessions wait for HBM.} A restoring tier moves a session's KV back into HBM
before the session runs (Fig.~\ref{fig:swap}a). For parked KV this costs little: restoring a resumed session's
7.5\,GB takes 0.53\,s from a Gen5 drive and 5\,ms from HBF at its 1.5\,TB/s interface, at most 1.7\% of a run, and host DRAM, a Gen5
drive and HBF finish within 2.3\% of each other at 32 to 256 sessions per generation (Table~\ref{tab:far}). A drive
thus swaps as well as HBF, yet swap alone does not end RSI's wait for HBM. First, only part of the KV is parked: under
load, ready sessions hold 95--99\% of it (\S\ref{sec:char-load}), and swap must keep all ready KV in HBM. Second, a
session switches between ready and parked dozens of times, mostly after under a second (Fig.~\ref{fig:switch}a), so
the KV that swap moves out is needed in HBM again almost at once. Ready sessions therefore queue for HBM, and a faster
tier shortens each restore but not the queue: with HBF or a drive, a call waits 11--12\,s for room in HBM at 32
sessions and 95--97\,s at 128. Tiering by access keeps
the hot set decode reads in HBM~\cite{baek_hbf}; here the hot set, the ready KV, alone exceeds HBM.

\begin{table}[t]
  \centering\footnotesize
  \caption{Run time with agentic KV beyond HBM on one far memory, by sessions per generation, as a multiple of NVMe
    swap's 174, 282, 490 and 940\,min.}
  \label{tab:far}
  \label{tab:parking}
  \label{tab:readinplace}
  \def\good#1{\textcolor{oursblue}{\textbf{#1}}}%
  \def\bad#1{\textcolor{black!50}{#1}}%
  \def\group#1{\rowcolor{black!6}\multicolumn{7}{@{\hspace{3pt}}l}{\textit{#1}}}%
  \setlength{\tabcolsep}{0pt}
  \begin{tabular*}{\columnwidth}{@{\extracolsep{\fill}\hspace{3pt}}lrrrrrl@{\hspace{3pt}}}
    \toprule
    \textbf{Far memory}           & \textbf{Read BW}  & \textbf{32} & \textbf{64} & \textbf{128} & \textbf{256} & \textbf{Verdict} \\
    \midrule
    None (HBM only)               & 2.8\,TB/s  & 2.25 & 2.87 & 3.49 & 3.69 & \bad{Too small} \\
    \group{Swap: parked KV moves into HBM before its session runs} \\
    Host DRAM~\cite{pcie5}                 & 25\,GB/s   & 0.99 & 1.00 & 0.99 & 0.99 & \bad{No gain} \\
    NVMe SSD~\cite{micron9550}             & 14\,GB/s   & 1.00 & 1.00 & 1.00 & 1.00 & \good{Fits} \\
    HBF~\cite{ocphbf}                      & 1.5\,TB/s  & 0.99 & 0.99 & 0.98 & 0.98 & \bad{No gain} \\
    \group{In place, unbounded: the GPU reads ready KV where it lies} \\
    Host DRAM                     & 64\,GB/s          & 7.0 & 15 & 22 & 25 & \bad{Too slow} \\
    NVMe SSD                      & 14\,GB/s          & 32 & 66 & 100 & 114 & \bad{Too slow} \\
    HBF                           & 467\,GB/s         & 1.29 & 1.16 & 0.83 & 0.50 & \good{Fits} \\
    \bottomrule
  \end{tabular*}
\end{table}

\noindent \textbf{Unbounded in-place reads lengthen every step.} Read in place (Fig.~\ref{fig:swap}b), KV beyond HBM
joins the step in which its session runs, so no session waits for room, but the far memory's bandwidth enters the
step. Reading 1\,GB of KV takes 0.36\,ms from HBM at the 2.8\,TB/s that paged decode attention reaches, 2.1\,ms from
one HBF site at 467\,GB/s, 15.6\,ms from host DRAM over PCIe and 71\,ms from an NVMe drive. An HBF site holds 22\x the capacity of an H200 HBM stack. Read in place whenever HBM is full, host DRAM takes 7--25\x and a drive 32--114\x as long as NVMe parking,
longer even than HBM alone
(Table~\ref{tab:far}). HBM and the far memory have separate channels, so the GPU reads both tiers concurrently, and a far memory of
bandwidth $B_F$ adds no time while a step reads at most $B_F\,t_H$ from it, $t_H$ being the step's HBM read time: a
49\,ms step over 99\,GB of KV and 34\,GB of weights hides 20\,GB from one HBF site, or 8.7\,GB at 202\,GB/s, and every
further gigabyte lengthens the step for every request in it. Reading the overflow from HBF whenever HBM is full thus
takes 30 and 17\% longer than parking at 32 and 64 sessions and pays off only from 128 on
(Table~\ref{tab:far}). Once ending sessions free HBM, flash reads waste time and energy.

\obsbox{2}{On-package in-place reads end the wait for HBM, if they hide behind HBM reads and stop once HBM frees.}

\begin{figure*}[t]
  \centering  \scalebox{0.92}{\input{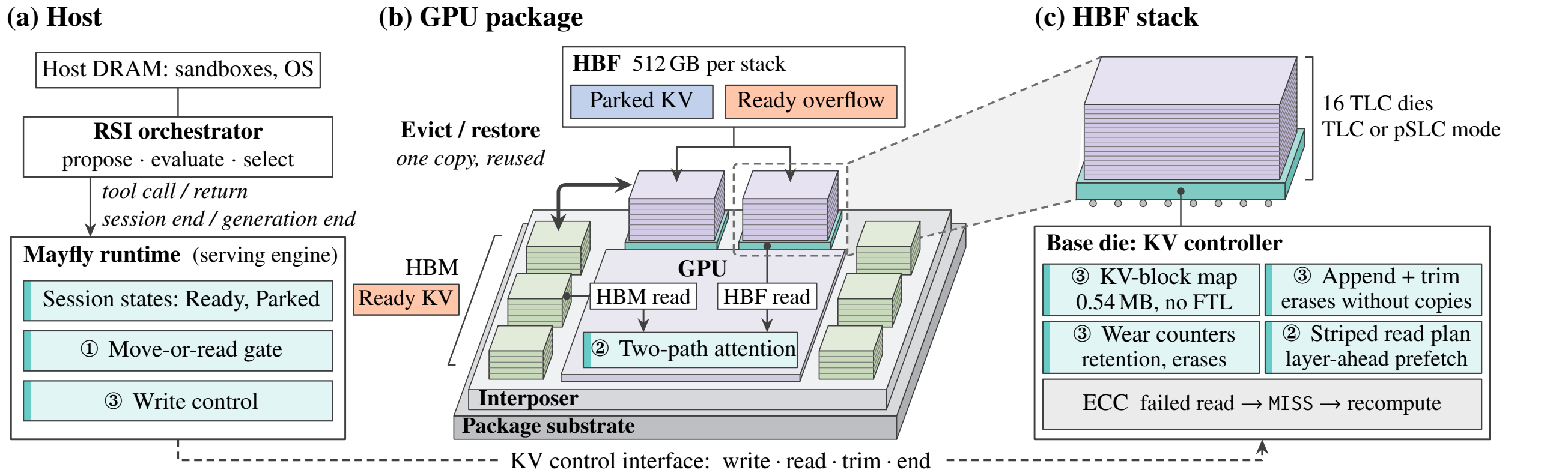}}
  \caption{\sys architecture: (a)~host, (b)~GPU package, (c)~HBF stack. Teal: added by \sys;
    \ding{172}--\ding{174}: the three mechanisms. A stack holds 512\,GB in TLC mode, or a third of it in pSLC mode at four times the P/E cycles (30K against 7.5K).}
\label{fig:hardware}
\end{figure*}

\subsection{Tiering by State or Timer Costs Writes}
\label{sec:char-life}

Flash pays off only for data read far more often than it is written~\cite{hbfsucks}. A stack's life falls with the
host's writes and with garbage collection's copies (\S\ref{sec:bg-hbf}), and two common policies raise both:
tiering by the session's state and retention for a fixed time.

\noindent \textbf{Tiering by state moves KV at every switch.} Parked and ready KV are the same blocks at different
moments. A real DGM session takes 39 turns and so switches state about 80 times, mostly briefly
(Fig.~\ref{fig:switch}): 88\% of parked intervals end within a second, while the 5\% over a minute, mostly test runs,
hold 93\% of the parked KV-time. A tier chosen by state therefore moves a block shortly after placing it. Demoting
every parked block to an NVMe drive below HBF saves no time, as most demoted blocks return within a second
(the run takes 0.4--3.4\% longer). It makes HBF absorb 13--16\x the writes, 170--280 against 11--21 drive
writes per day (DWPD). HBF that only parks finishes with the drive while absorbing
18\,DWPD at 32 sessions per generation, against 2.4 for the 7.5\x larger drive. Reads far outnumber switches: a
session's blocks are read at every step that schedules it. A
tier that takes parked KV and returns it writes each block about once per turn; a copy that stays valid across
switches is written once.

\noindent \textbf{Timed retention keeps dead KV.} An agent server cannot know which response is a session's last: in
TraceLab, users reply after more than an hour in 3.4\% of the gaps. It therefore keeps KV for a fixed time
(Fig.~\ref{fig:life}); with the 5\,min of hosted prompt caches, 13\% of the KV kept for users belongs to ended
sessions. A fixed time suits RSI no better: 5\,min expires sessions waiting on long tool
calls, which recompute 9--17\% more prefill tokens, and 30\,min keeps up to 1.5\,TB live at 128 sessions per
generation. A DGM session's KV, 9.9\,GB at the median, dies at a session end that the orchestrator announces. A flash controller that is not told copies such
dead but valid KV in every garbage collection that reaches it, and the WAF reaches 1.13--1.90 at 128 sessions per
generation.

\obsbox{3}{Reads far outnumber state switches, and deaths are announced; tiering by state or timer wastes writes.}

\subsection{Design Requirements}
\label{sec:char-motiv}

The three costs share one cause: each approach lets the switch between ready and parked decide more than it
determines. Restoring ties where KV resides to the switch, unbounded in-place reads tie what a step reads to it, and
tiering by state ties the flash copy to it, inferring the block's end from idleness. Our key insight is that a switch
decides only whether a block is read, so its flash copy stays valid (\need{1}); free HBM and the step's bandwidth decide where it is read from, so a step reads ready KV in place within a bound (\need{2}) and KV returns to HBM
as sessions end (\need{3}); and the session's end, which the orchestrator announces, decides when its copy dies (\need{4}). Decided
apart, they let ready sessions run without waiting for HBM, keep each step's flash reads within a bound, and write each
block once and reclaim it when it dies.

\section{\sys Design}
\label{sec:design}

\sys adds a serving runtime on the host, a two-path attention kernel on the GPU and a KV controller on the HBF base
die (Fig.~\ref{fig:hardware}), and meets the design requirements of \S\ref{sec:char-motiv} with three mechanisms around one
backing copy per block, which stays valid across state switches (\need{1}).
\ding{172}~\emph{Move or read.} Before each step, a gate in the runtime decides which ready requests run and whether
the KV each holds beyond HBM is restored or read in place (\need{2}), and KV returns to HBM as sessions end (\need{3}).
\ding{173}~\emph{Read two tiers within a step.} The kernel, fed by reads striped on the base die, computes exact
attention over a request's KV in both tiers.
\ding{174}~\emph{Write control.} The runtime writes only live blocks that lack a copy, and the controller invalidates
a session's KV when the orchestrator announces its end (\need{4}).
Most prior HBF tiers key these decisions to hotness, context length, KV type or age, or fix them when KV is allocated~\cite{lachesis}; \sys re-decides each at the RSI event that settles it (\S\ref{sec:char-motiv}): a state switch keeps the block's flash copy, free HBM and the step's bandwidth decide where ready KV is read, and an announced session end retires its KV.
\sys adds HBF sites beside the HBM stacks and leaves the GPU cores and the HBM and NAND dies unchanged; the orchestrator keeps its control flow and adds two notifications, where
it already ends a session and a generation. Table~\ref{tab:prior} compares prior HBF designs: none meets more than one
requirement in full, and none returns KV to HBM as sessions end.

\begin{table}[t]
  \centering\footnotesize
  \definecolor{cyesgreen}{HTML}{2E8B3D}\definecolor{cnored}{HTML}{C9302C}\definecolor{cpartamber}{HTML}{D98E04}%
  \def\cyes{\textcolor{cyesgreen}{\ding{51}}}%
  \def\cno{\textcolor{cnored}{\ding{55}}}%
  \def\cpart{\tikz[baseline=-0.6ex]{\draw[cpartamber,line width=0.5pt] (0,0) circle (0.62ex);
    \fill[cpartamber] (0,0.62ex) arc (90:270:0.62ex) -- cycle;}}%
  \caption{Prior HBF designs against the design requirements. \protect\cyes~yes, \protect\cpart~partly, \protect\cno~no or not described,
    --~not applicable.}
  \label{tab:prior}
  \setlength{\tabcolsep}{0pt}
  \begin{tabular*}{\columnwidth}{@{\extracolsep{\fill}\hspace{3pt}}lcccccc@{\hspace{3pt}}}
    \toprule
                    &                &                & \need{1}      & \need{2}       & \need{3}        & \need{4} \\
                    & \textbf{In}    & \textbf{Exact} & \textbf{Copy} & \textbf{Step}  & \textbf{Back to} & \textbf{Freed} \\
    \textbf{Design} & \textbf{place} & \textbf{attn.} & \textbf{kept} & \textbf{bound} & \textbf{HBM}    & \textbf{at end} \\
    \midrule
    Baek et al.~\cite{baek_hbf}  & \cno   & \cyes & \cyes & --    & --   & \cno   \\
    SETA~\cite{seta}             & \cyes  & \cyes & \cno  & \cno  & \cno & \cno   \\
    FlashAccel~\cite{flashaccel} & \cyes  & \cyes & \cyes & \cno  & \cno & \cno   \\
    HBFlex~\cite{hbflex}         & \cyes  & \cyes & --    & \cno  & --   & \cpart \\
    Yu et al.~\cite{yu_hbf}      & \cyes  & \cyes & --    & \cno  & \cno & \cno   \\
    SPLASH~\cite{splash}         & \cyes  & \cno  & --    & \cyes & \cno & \cpart \\
    Lachesis~\cite{lachesis}     & \cyes  & \cyes & --    & \cno  & \cno & \cyes  \\
    \midrule
    \textbf{\sys}                & \cyes  & \cyes & \cyes & \cyes & \cyes & \cyes \\
    \bottomrule
  \end{tabular*}
\end{table}

\begin{figure*}[t]
  \centering
  \resizebox{\linewidth}{!}{
\begin{tikzpicture}[x=1cm, y=0.9cm, line join=miter,
    every node/.style={font=\fontfamily{ptm}\fontsize{8}{9.4}\selectfont, inner sep=0pt, outer sep=0pt},
    mf blk/.style={draw=mfline, line width=0.5pt, inner sep=1.5pt, align=center},
    mf old/.style={mf blk, fill=white},
    mf ready/.style={mf blk, fill=mfready},
    mf park/.style={mf blk, fill=mfpark},
    mf held/.style={mf blk, fill=mfheld},
    mf abs/.style={draw=mfline!75, line width=0.4pt, dash pattern=on 1.7pt off 1.3pt, inner sep=1.5pt},
    mf box/.style={minimum height=0.4cm},
    mf ttl/.style={anchor=north west, font=\fontfamily{ptm}\bfseries\fontsize{9}{10.8}\selectfont},
    mf arr/.style={line width=0.6pt, draw=mfline, -{Stealth[length=1.6mm,width=1.3mm]}},
    mf line/.style={line width=0.6pt, draw=mfline},
    mf guide/.style={line width=0.35pt, draw=mfline!55, dash pattern=on 1.4pt off 1.4pt},
  ]
  \definecolor{mfline}{HTML}{404040}
  \definecolor{mfcard}{HTML}{F2F2F3}
  \definecolor{mfrule}{HTML}{B4B8BC}
  \definecolor{mfnew}{HTML}{E1F5F4}
  \definecolor{mfstripe}{HTML}{68CCC6}
  \definecolor{mfready}{HTML}{FFC6A9}
  \definecolor{mfpark}{HTML}{C2D1EB}
  \definecolor{mfheld}{HTML}{D9D9D9}
  \definecolor{hbmT}{HTML}{E2EBDA}
  \definecolor{hbfT}{HTML}{E7E2F1}
  \def\mfS{\fontsize{7}{8.2}\selectfont\color{black}}%
  \def\kv{{\itshape KV}}%
  \def\card#1#2#3#4{\fill[mfcard, rounded corners=4pt] (#1,#2) rectangle (#3,#4);}%
  \def\mfmod#1#2#3#4#5{%
    \fill[mfnew] (#1,#2-#4/2) rectangle ++(#3,#4);
    \fill[mfstripe] (#1,#2-#4/2) rectangle ++(0.08,#4);
    \draw[draw=mfline, line width=0.45pt] (#1,#2-#4/2) rectangle ++(#3,#4);
    \node[align=center] at (#1+#3/2+0.04,#2) {#5};}%
  \def\seg#1#2#3#4#5{\node[#3, minimum width=#2cm-#1cm, minimum height=0.46cm, anchor=south west] at (#1,#4) {#5};}%

  \card{0}{4.42}{17.7}{7.9}
  \node[mf ttl] at (0.18,7.8) {(a) Block lifecycle};
  \node[anchor=north east] at (17.5,7.74) {%
    \tikz[baseline=-0.6ex]\draw[line width=0.4pt, draw=mfline, fill=mfready] (0,-0.09) rectangle (0.4,0.11);~read\qquad
    \tikz[baseline=-0.6ex]\draw[line width=0.4pt, draw=mfline, fill=mfheld] (0,-0.09) rectangle (0.4,0.11);~stored\qquad
    \tikz[baseline=-0.6ex]\draw[line width=0.4pt, draw=mfline!75, dash pattern=on 1.7pt off 1.3pt] (0,-0.09) rectangle (0.4,0.11);~absent};
  \node[anchor=east] at (1.25,6.57) {\textbf{Session}};
  \node[anchor=east] at (1.25,5.71) {\textbf{HBM}};
  \node[anchor=east] at (1.25,4.85) {\textbf{HBF}};
  \foreach \x in {3.9, 6.5, 9.4, 12.3, 15.1} \draw[mf guide] (\x,6.32) -- (\x,4.5);
  \seg{1.4}{3.9}{mf ready}{6.32}{Ready}
  \seg{3.9}{6.5}{mf park}{6.32}{Parked}
  \seg{6.5}{12.3}{mf ready}{6.32}{Ready}
  \seg{12.3}{15.1}{mf park}{6.32}{Parked}
  \seg{15.1}{17.5}{mf old}{6.32}{Ended}
  \foreach \x/\t in {3.9/Tool call, 6.5/{Tool return; HBM full}, 9.4/HBM has room, 12.3/Tool call, 15.1/Session end} {
    \draw[mf arr] (\x,7.04) -- (\x,6.84);
    \node[anchor=south] at (\x,7.08) {\mfS\itshape \t};
  }
  \seg{1.65}{3.65}{mf ready}{5.46}{\kv\ (read)}
  \seg{4.75}{6.25}{mf abs}{5.46}{}
  \seg{6.9}{8.9}{mf abs}{5.46}{}
  \seg{9.95}{12.0}{mf ready}{5.46}{\kv\ (read)}
  \seg{12.65}{14.75}{mf abs}{5.46}{}
  \seg{15.45}{17.25}{mf abs}{5.46}{}
  \seg{1.65}{3.65}{mf abs}{4.6}{}
  \seg{4.3}{6.25}{mf held}{4.6}{\kv\ (stored)}
  \seg{6.85}{9.15}{mf ready}{4.6}{\kv\ (read in place)}
  \seg{9.95}{14.75}{mf held}{4.6}{\kv\ (stored; copy retained)}
  \seg{15.45}{17.25}{mf old}{4.6}{\kv\ (trimmed)}
  \draw[line width=0.35pt, draw=mfline!70] (15.45,4.6) -- (17.25,5.111) (15.45,5.111) -- (17.25,4.6);
  \draw[mf arr] (3.05,5.46) -- (4.3,4.9);
  \node[anchor=west] at (3.72,5.3) {\mfS\textbf{\itshape Write once}\, (HBM pressure)};
  \draw[mf arr] (9.15,4.98) -- (9.95,5.56);
  \node[anchor=east] at (9.22,5.3) {\mfS\bfseries\itshape Restore};
  \node[anchor=center] at (13.55,5.285) {\mfS\bfseries\itshape No write: copy reused};

  \card{0}{0}{5.83}{4.3}
  \node[mf ttl] at (0.18,4.2) {(b) Move-or-read gate};
  \node[mf old, mf box, minimum width=2.6cm] (b0) at (1.8,3.5) {Ready request};
  \mfmod{0.5}{2.76}{2.6}{0.42}{Fits in HBM?}
  \mfmod{0.5}{1.74}{2.6}{0.8}{Within step budget?\\[1pt]\mfS $t_1-t_0\le t_0/k$}
  \node[mf old, mf box, minimum width=2.6cm] (b3) at (1.8,0.62) {Defer to next step};
  \node[mf ready, mf box, minimum width=2.0cm] (b1) at (4.55,2.76) {Restore to HBM};
  \node[mf ready, mf box, minimum width=2.0cm] (b2) at (4.55,1.74) {Read in place};
  \node[mf old, mf box, minimum width=2.0cm] (b4) at (4.55,0.62) {\textbf{Admit}};
  \draw[mf arr] (b0.south) -- (1.8,2.99);
  \draw[mf arr] (1.8,2.53) -- (1.8,2.18);
  \node[anchor=west] at (1.9,2.36) {\mfS No};
  \draw[mf arr] (1.8,1.3) -- (b3.north);
  \node[anchor=west] at (1.9,1.08) {\mfS No};
  \draw[mf arr] (3.1,2.76) -- (b1.west);
  \node[anchor=south] at (3.32,2.82) {\mfS Yes};
  \draw[mf arr] (3.1,1.74) -- (b2.west);
  \node[anchor=south] at (3.32,1.8) {\mfS Yes};
  \draw[mf line] (b1.east) -- (5.7,2.76) -- (5.7,1.74);
  \draw[mf line] (b2.east) -- (5.7,1.74);
  \draw[mf arr] (5.7,1.74) |- (b4.east);
  \draw[mf arr] (b3.west) -- (0.18,0.62) |- (b0.west);

  \card{5.97}{0}{11.73}{4.3}
  \node[mf ttl] at (6.15,4.2) {(c) Two-path attention};
  \node[anchor=west] at (6.85,3.7) {One request's \kv};
  \foreach \i in {0,1,2} \node[mf blk, fill=hbmT, minimum width=0.8cm, minimum height=0.36cm, anchor=south west] at (6.85+\i*0.8,3.08) {HBM};
  \foreach \i in {3,4} \node[mf blk, fill=hbfT, minimum width=0.8cm, minimum height=0.36cm, anchor=south west] at (6.85+\i*0.8,3.08) {HBF};
  \mfmod{6.85}{2.58}{2.3}{0.42}{HBM path}
  \mfmod{9.35}{2.58}{1.5}{0.42}{HBF path}
  \draw[mf arr] (8.0,3.08) -- (8.0,2.82);
  \draw[mf arr] (10.1,3.08) -- (10.1,2.82);
  \mfmod{6.85}{1.2}{4.0}{0.42}{Log-sum-exp merge}
  \draw[mf arr] (7.12,2.35) -- (7.12,1.44);
  \draw[mf arr] (9.62,2.35) -- (9.62,1.44);
  \node[anchor=west, align=left, font=\fontfamily{ptm}\fontsize{7}{8}\selectfont] at (7.24,1.9) {Partial output +\\softmax statistics};
  \node[anchor=west, align=left, font=\fontfamily{ptm}\fontsize{7}{8}\selectfont] at (9.74,1.9) {Partial output +\\softmax statistics};
  \node[mf old, mf box, minimum width=4.0cm] (c9) at (8.85,0.42) {Exact attention};
  \draw[mf arr] (8.85,0.96) -- (c9.north);

  \card{11.87}{0}{17.7}{4.3}
  \node[mf ttl] at (12.05,4.2) {(d) Write control};
  \node[mf old, mf box, minimum width=2.4cm] (d0) at (13.3,3.6) {Evicted block};
  \mfmod{12.1}{2.92}{2.4}{0.42}{Session ended?}
  \mfmod{12.1}{2.26}{2.4}{0.42}{Valid HBF copy?}
  \node[mf old, mf box, minimum width=2.4cm] (d3) at (13.3,1.6) {\textbf{Write to HBF once}};
  \node[mf old, mf box, minimum width=2.3cm] (d1) at (16.4,2.92) {Drop: no write};
  \node[mf old, mf box, minimum width=2.3cm] (d2) at (16.4,2.26) {Reuse: no write};
  \draw[mf arr] (d0.south) -- (13.3,3.16);
  \draw[mf arr] (13.3,2.69) -- (13.3,2.5);
  \node[anchor=west] at (13.4,2.6) {\mfS No};
  \draw[mf arr] (13.3,2.03) -- (d3.north);
  \node[anchor=west] at (13.4,1.93) {\mfS No};
  \draw[mf arr] (14.5,2.92) -- (d1.west);
  \node[anchor=south] at (14.87,2.98) {\mfS Yes};
  \draw[mf arr] (14.5,2.26) -- (d2.west);
  \node[anchor=south] at (14.87,2.32) {\mfS Yes};
  \draw[draw=mfrule, line width=0.5pt] (12.1,1.23) -- (17.55,1.23);
  \node[anchor=east] at (13.42,0.9) {\mfS\bfseries No trim};
  \node[anchor=east] at (13.42,0.32) {\mfS\bfseries After trim};
  \foreach \c/\d in {0/0,1/1,2/1,3/0,4/1,5/0} {
    \ifnum\d=1
      \draw[line width=0.35pt, draw=mfline, fill=white] (13.55+\c*0.3,0.76) rectangle ++(0.3,0.28);
      \draw[line width=0.35pt, draw=mfline] (13.55+\c*0.3,0.76) -- ++(0.3,0.28) (13.55+\c*0.3,1.04) -- ++(0.3,-0.28);
    \else
      \draw[line width=0.35pt, draw=mfline, fill=hbfT] (13.55+\c*0.3,0.76) rectangle ++(0.3,0.28);
    \fi}
  \foreach \c in {0,...,5}
    \draw[line width=0.35pt, draw=mfline, fill=white] (13.55+\c*0.3,0.18) rectangle ++(0.3,0.28);
  \draw[mf arr] (15.45,0.9) -- (15.82,0.9);
  \draw[mf arr] (15.45,0.32) -- (15.82,0.32);
  \node[anchor=west, align=left, font=\fontfamily{ptm}\fontsize{7}{7.6}\selectfont] at (15.9,0.9) {Dead but valid:\\GC copies};
  \node[anchor=west, align=left, font=\fontfamily{ptm}\fontsize{7}{7.6}\selectfont] at (15.9,0.32) {All dead:\\erase, no copy};
\end{tikzpicture}%
}
  \caption{\sys at run time: (a)~a full KV block across session states; (b)~the gate, for each ready request at every
    step; (c)~one step's attention over HBM and HBF; (d)~eviction writes and an erase unit without and with trims.}
  \label{fig:workflow}
\end{figure*}
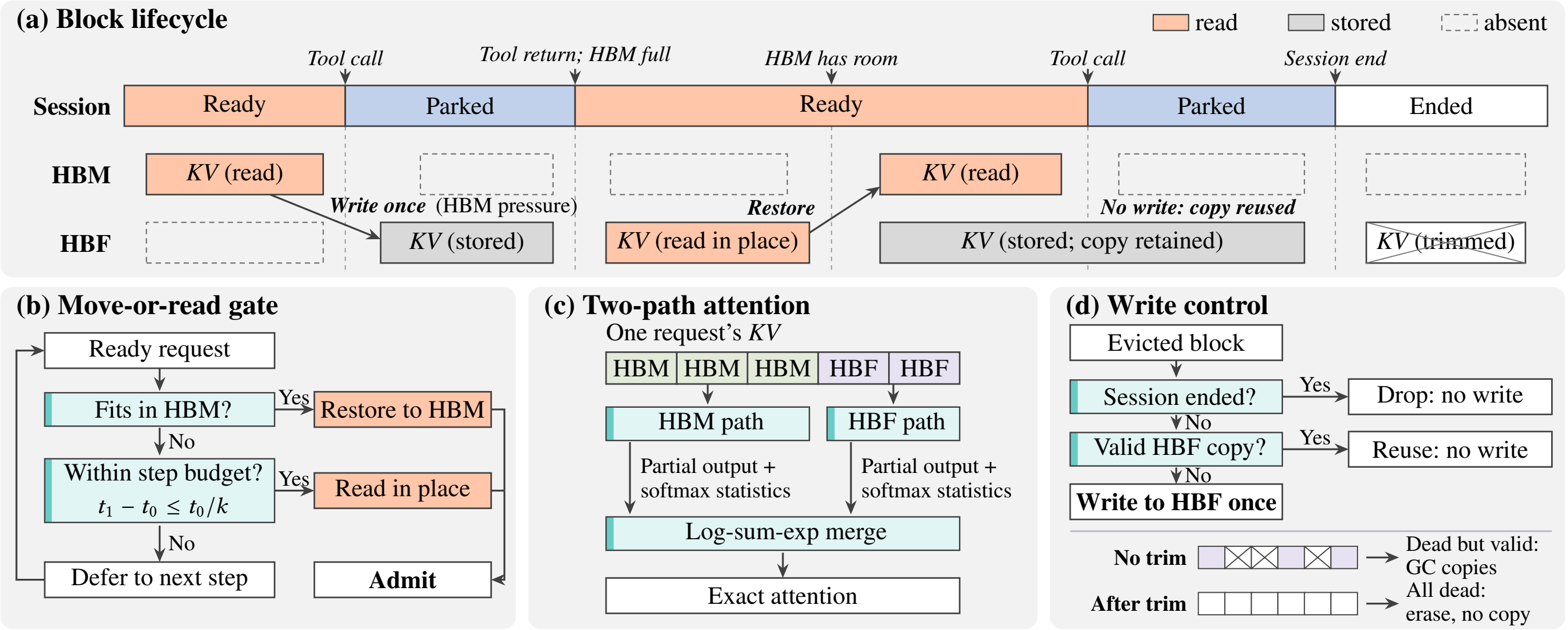

\subsection{One Backing Copy, Two States}
\label{sec:onecopy}
\label{sec:hints}

KV lives in vLLM's 16-token blocks, 4\,MiB each, and \sys tracks three things apart (Fig.~\ref{fig:workflow}a): a
session's state, ready or parked until it ends, held only in the runtime; where each block resides, in HBM or on HBF; and
whether the block has a valid backing copy on HBF. A block is immutable once its 16 tokens are filled, and only full
blocks leave HBM; the block that a request is still filling stays there. The copy follows one rule: it is written when
its block first leaves HBM and stays valid until the block's last session ends. Restoring the block to HBM keeps the copy, and
evicting it again writes nothing. A state switch thus
moves no KV by itself. When a session calls a tool, it turns parked, and its blocks leave HBM only under pressure; when
the tool returns, it turns ready, and the gate restores its blocks or lets the GPU read them in place
(\S\ref{sec:gate}). HBM keeps, in this order, the task prefixes that live sessions share, ready KV and parked KV, each
context's tail leaving first. Parked KV stays on HBF: a block demoted to a drive is
soon written to HBF again, and one tier spares a drive per GPU and a second placement decision.

\noindent \textbf{Signals and commands.}
The runtime needs three signals from the loop. Tool calls and returns, which the serving engine already sees as one
request ending and the next arriving, mark a session parked or ready; nothing predicts when a tool returns, since ready
KV is read wherever it lies. The orchestrator adds the other two: \texttt{session\_end(s)} trims a session's private
blocks, and \texttt{gen\_end(g)}, sent once all of a generation's sessions have ended, trims the prefixes they share and
backs up lost session ends. The runtime drives the controller with four commands (Fig.~\ref{fig:hardware}):
\texttt{write(block, session, gen, class)}, with class \emph{shared} for task prefixes and \emph{session} for private KV;
\texttt{read(block)}, which returns the data or \texttt{MISS}; \texttt{trim(session)}; and \texttt{end(gen)}. The signals
affect only performance and wear: a lost \texttt{session\_end} leaves dead KV in flash until its generation ends or its
retention target expires.

\subsection{Move or Read: A Per-Step Gate}
\label{sec:gate}
\label{sec:runtime}

Before each step, the gate chooses the requests that run and where each one's KV is read (Fig.~\ref{fig:workflow}b).
It starts from the bound of \S\ref{sec:whyhbf}: HBF reads that finish within the step's HBM reads add no time. It
estimates the time of a step over a set $A$ of requests as $T(A)=c_0+\max(t_H,t_F)$. Here $c_0$ covers scheduling,
launch and sampling; $t_H=t_W+K_H/B_H$ reads the weights, in $t_W$, and the set's KV in HBM, $K_H$, at HBM's bandwidth
$B_H$; and $t_F=K_F/B_F+K_W/B_W$ reads the set's KV on HBF, $K_F$, and writes the KV leaving HBM, $K_W$. The HBM and HBF
paths run in parallel, hence the maximum, whereas HBF's reads and writes share its half-duplex interface and add up.
$B_F$ and $B_W$ sum over the package's HBF stacks, across which the runtime spreads each session's blocks evenly.
The gate builds the step's set $A$, with $k=|A|$, in the serving engine's order: running requests oldest first, then
waiting ones by arrival. It admits the first request outright, and any request whose KV fits in HBM together with that
of $A$. Any other request $r$ would read in place, and the gate admits it only if it does not lower the set's estimated
token throughput. As each request decodes one token per step, the test is $(k+1)/t_1\ge k/t_0$, with $t_0=T(A)$ and
$t_1=T(A\cup\{r\})$, that is, $t_1-t_0\le t_0/k$. The test is looser than the bound: $r$ may outlast the step's HBM
reads when its own token pays for the added time. An unadmitted request
skips the step, keeps its place and leaves its KV on HBF, and a waiting request is admitted only if no running
one skipped, so new sessions do not overtake skipped ones. The test protects throughput rather than any one request's latency.

\noindent \textbf{Returning to HBM.}
\label{sec:return}
KV returns to HBM in the step that reads it. After each step, the runtime reassigns HBM in the step's order: the KV of
its requests comes first, the oldest request first, and parked KV yields. A block that the step read from HBF and that
now has room stays in HBM, at one HBM write and no extra HBF read; the block table switches to the HBM copy after the
step, once every layer of the block has been read, and the HBF copy stays valid. A resumed session whose KV fits is
restored this way in its first step. A generation's sessions end one by one and free HBM, while its slowest sessions
keep decoding and set its run time (\S\ref{sec:char-time}), so each step shrinks the ready KV beyond HBM; once it fits,
every request runs without HBF reads, and HBF falls idle.

\subsection{Reading Two Tiers Within a Step}
\label{sec:readpath}

\sys assumes, as SPLASH~\cite{splash}, HBFlex~\cite{hbflex} and FlashAccel~\cite{flashaccel} do, that the GPU addresses an HBF stack as it addresses the HBM stacks beside it
(Fig.~\ref{fig:hardware}b): the stack's blocks are mapped into the GPU's physical address space, and the attention
kernel loads them like HBM blocks. An entry of vLLM's block table may point into either tier, and the kernel splits a
request's KV by tier (Fig.~\ref{fig:workflow}c), as Flash-Decoding splits a long context across thread
blocks~\cite{flashdecoding}: one group of thread blocks attends over the HBM blocks, another over the HBF blocks, each
yields a partial output with its softmax maximum and sum, and a log-sum-exp reduction merges the two into exact
attention over the whole context, so no KV is dropped to save reads. The two groups
run concurrently, so $T$ takes the longer of the two paths, and the merge adds one reduction per layer.

\noindent \textbf{Striped read plan.}
Before a step starts, the runtime sends the controller the step's \texttt{read} commands, one per block that its
requests hold on HBF. Every address is thus known before it is needed, and nothing is predicted, unlike MoE expert
prefetch on HBF~\cite{hypaic}. The controller stripes each block's pages over the stack's dies and planes, so that the
pages one layer needs keep every plane busy, and issues the NAND reads of layer $\ell{+}1$ while the GPU runs layer
$\ell$. A buffer of twice the transfer's latency--bandwidth product hides the move from the dies' page registers to the
GPU, so the bytes the array returns per page read, not one read's latency, limit the read: $D$ dies of $P$ planes
return $D\,P$ pages of 4\,KiB~\cite{ocphbf} per read time, 1.68\,TB/s for 16 dies of 512 planes at 20\,$\mu$s, 3.6\x what a site needs. The rated interface implies such
parallelism: 16 dies reach the 1.5\,TB/s grade at a 20\,$\mu$s page read only with about 460 planes of 4\,KiB each.

\subsection{Write Control}
\label{sec:writectl}

A conventional tier, a victim cache, writes every block evicted from HBM. \sys decides at each eviction, in this order
(Fig.~\ref{fig:workflow}d): a block that no live session references is dropped, and such blocks leave HBM first; a block
with a valid backing copy is freed in HBM without a write; and only a live block without a copy is written, once.
Unlike reuse-proven admission~\cite{saria}, this needs no evidence of reuse, because a live session's KV is read when
the session runs. Death, not garbage collection, then frees flash. On \texttt{session\_end}, the runtime trims the
session's private blocks, and a shared prefix dies with its last user, at the latest at \texttt{gen\_end}. A trim only
marks blocks dead: garbage collection no longer copies them, and an erase unit is erased without copies once all its
blocks are dead. A session's blocks are written close together and die together, so in one append stream most units
empty before they are reclaimed, without per-session placement. Against lost session ends, the
controller keeps open erase units per generation and class, and \texttt{end(gen)} erases them whole, shared prefixes
included.

\subsection{KV-Native HBF Controller}
\label{sec:controller}

A conventional flash controller serves 4\,KiB writes that must survive power loss and may be read years later, so its page
map for a 512\,GB stack needs about 500\,MB of journaled DRAM, and it learns that data is dead only when the host
overwrites or discards it. RSI KV is written in large blocks, can be regenerated by prefill, announces its death and is
read within hours, and the \sys controller exploits all four properties with the parts shown in
Fig.~\ref{fig:hardware}c.

\noindent \textbf{Map and allocation.}
The controller maps whole KV blocks, not pages, and appends them to open erase units. An erase unit is a superblock, the
NAND blocks that a stripe erases together, of 64\,MB to 1\,GB. A stack's 122K blocks and its table
of the smallest units take 0.54\,MB of SRAM, under 0.1\% of a 121--176\,mm$^2$ base die, where the 9.6\,MB read buffer
takes 1.1--1.6\% and prior HBF designs place 32--40\,MB of buffer SRAM~\cite{flashaccel,hbflex}. KV can be recomputed,
so the controller keeps no journal and, after a power loss, reports every block missing, to be recomputed on demand.

\noindent \textbf{Retention.}
NAND is specified to retain data for three months to a year at end of life, while RSI KV is read within
94\,min of its write. The controller records each erase unit's program time and retention target; once a unit outlives
its target, its blocks read as \texttt{MISS} and are recomputed, never refreshed, since refreshing KV that may never be
read adds writes. A target below the specification's 24\,h of powered retention~\cite{ocphbf} can trade
retention for endurance.

\noindent \textbf{Missing blocks and error correction.}
A block reads as \texttt{MISS} when its unit has outlived its retention target, when a codeword fails its CRC, or after a
power loss. The kernel flags the block's request, and the step's other requests commit their tokens. The runtime, which
keeps every block's tokens, recomputes the block into HBM by a prefill of its 16 tokens over the KV before it, earlier
missing blocks first, trims the failed copy and reruns the request's step with exact attention; a miss whose earlier
blocks are present thus adds one block's prefill and one step to its own request. Any HBF base die needs error
correction~\cite{ocphbf}, which prior HBF studies charge nothing~\cite{hbfsucks,hbfsim} or omit~\cite{flashaccel,hbflex}.
As a codeword with more errors than the code corrects costs a recompute and no data, the KV's read volume, not a
storage-grade UBER, sets the target. As
cells wear, more blocks read as \texttt{MISS} and worn units retire, so the tier loses capacity instead of data.

\section{Evaluation}
\label{sec:eval}

\subsection{Experimental Setup}
\label{sec:method}
\label{sec:method-models}

\noindent \textbf{Workloads.} We adopt three harness-RSI systems. DGM~\cite{dgm} evolves a coding
agent on SWE-bench; our run, served by gpt-oss-120b~\cite{gptoss}, has five generations, 147 sessions and 7,008
turns, and a second run served by GPT-5 nano~\cite{gpt5nano} has 197 sessions and 6,814 turns. RRSI~\cite{rrsi}
refines an agent harness on EngDesign over a baseline and three rounds; our run, served by gpt-oss-120b, has 4,807
sessions and 24,740 turns. Harness$^2$~\cite{harness2}, an independent framework, refines an OpenCode agent's
harness; our run on its 25 LAB tasks, served by Qwen3-32B on an H200, has 664 sessions and 8,584 calls. Every replay
keeps each session's turns, contexts and tool times and serves them on Qwen3-32B with FP8 weights. From the
first DGM run's evaluation sessions we build the loop of Fig.~\ref{fig:burst}a: in each of three generations, $V$
variants attempt the same eight tasks, the $8V$ sessions start together and the next generation when the
slowest ends; extra attempts repeat a recorded one with KV of its own. $V{=}4$, 8, 16 and 32 give 32 to 256 sessions
per generation. The recorded run has ten tasks with more than one attempt; results use one sample of eight of them unless
marked as three- or ten-sample means, and speedups are ratios of run times. A discrete-event simulator re-implements vLLM's block pool, scheduler and preemption, with step time and energy fitted on H200 replays, one of which spills KV to host DRAM (within 0.5\%); on a held-out replay with a 20\,GB KV pool, it is within 0.9\% of the measured run time, 1.6\% of the energy and 5\% of the preemptions. A step that reads HBF takes the gate's $T$ (\S\ref{sec:gate}), with the compute time in its
maximum and $c_0$ fitted at 6.0\,ms on H200 replays. We check the HBF path in parts: an H200
two-path kernel over host memory, a cycle-approximate read pipeline and a gate with misestimated bandwidth.

\noindent \textbf{Hardware configuration.} The GPU is an H200 whose six HBM sites hold 141\,GB, 99\,GB of it for
KV. HBF sits on package sites added beside them, one 512\,GB TLC stack per 32 sessions, below the 550\,GB that the OCP specification targets.
A site reads 467\,GB/s, the rate at which an H200 HBM site serves paged decode attention and below the OCP grades of 1.5
and 3.1\,TB/s~\cite{ocphbf}, and writes a tenth of it~\cite{hbflex,seta}; we also run 202\,GB/s, a stack's rate under
a thermal limit~\cite{hbfsucks}. The specification leaves energy and array timing open, so we take 20 and
100\,pJ/bit for reads and writes and 20\,$\mu$s page reads from the high end of prior
estimates~\cite{flashaccel,hbfsucks,yu_hbf,hbfsim}. Every tier write and trim is replayed through a flash model
with greedy garbage collection, and we report wear: with wear spread evenly, each block spends DWPD $\times$ WAF
of its P/E cycles a day, with P/E budgets measured on 3D TLC~\cite{aero,star}.

\noindent \textbf{Baselines.} Every baseline keeps the six HBM sites and puts KV beyond HBM on one far memory: HBF;
host DRAM over PCIe, at 25\,GB/s as a parking tier and 64\,GB/s read in place (25\,pJ/bit); Grace LPDDR5X over
NVLink-C2C (480\,GB, 450\,GB/s, 8\,pJ/bit); or a Gen5 NVMe drive (3.84\,TB, 14\,GB/s reads and 10\,GB/s writes, 300
and 600\,pJ/bit). Parking on the drive or HBF (NVMe-Park, HBF-Park) admits a ready session only once its KV fits in HBM and has been restored there. HBF-Park gets \sys's HBF capacity
and write control, keeps a block's flash copy valid across resumes, and restores at the OCP interface's 1.5\,TB/s. Without a read bound, HBF-All reads all ready KV from HBF at every step and
HBF-Greedy reads ready KV in place whenever HBM is full. All HBF designs run on the same added sites, so every gain over an HBF
baseline comes at equal package resources.

\subsection{Overall Performance}
\label{sec:eval-inplace}
\label{sec:eval-dgm}

\begin{figure}[t]
  \centering
  \includegraphics[width=\columnwidth]{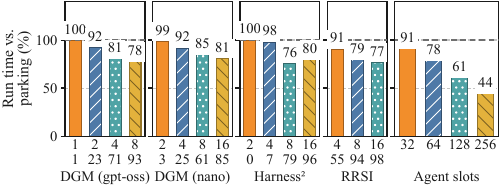}
  \caption{Run time of \sys in \% of HBF-Park's (whole runs per H200) and NVMe-Park's (agent slots). Labels: runs per H200 and, below them, time with ready KV over HBM (\%).}
  \label{fig:real}
\end{figure}

\sys's gain follows ready-KV overflow and grows with the load.

\noindent \textbf{Whole runs.} We replay whole runs, $R$ per H200 started 900\,s apart on two HBF sites (Fig.~\ref{fig:real}): the
DGM run of Fig.~\ref{fig:burst}, its GPT-5 nano counterpart, the RRSI run and the Harness$^2$ run. A DGM run keeps at most 10 sessions open, so with one run per H200 its
ready KV exceeds HBM 1.4\% of the time. While ready KV stays within HBM, all designs finish within 3.1\% of each other. Once it exceeds
HBM most of the time, \sys finishes 19 and 22\% sooner than HBF-Park with 4 and 8 DGM runs per H200, and 24 and 20\% sooner with 8 and 16 Harness$^2$ runs. The GPT-5 nano run, whose shorter sessions overflow HBM as
often only at twice the runs, finishes 15 and 19\% sooner with 8 and 16 runs. RRSI finishes 9, 21 and 23\% sooner
with 4, 8 and 16 runs. HBM alone takes up to 5.1\x as long.

\begin{figure}[t]
  \centering
  \includegraphics[width=\columnwidth]{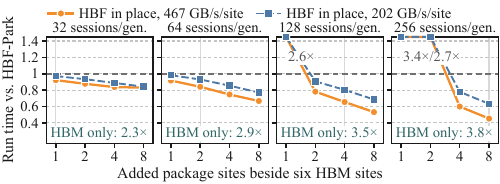}
  \caption{Run time over HBF-Park by HBF sites added, at 467 or 202\,GB/s each. Vertical line: one site per 32
    sessions; top labels: off-scale points.}
  \label{fig:inplace}
\end{figure}

\begin{table}[t]
  \centering\footnotesize
  \caption{Where the gain comes from, by sessions per generation. TB written: to HBF or NVMe-Park's drive. Bold: best.}
  \label{tab:gain}
  \definecolor{sysband}{HTML}{E1F5F4}%
  \setlength{\tabcolsep}{0pt}\setlength{\aboverulesep}{0pt}\setlength{\belowrulesep}{0pt}%
  \def\tup{\rule{0pt}{2.5ex}}\def\tdn{\rule[-1.0ex]{0pt}{0pt}}%
  \begin{tabularx}{\columnwidth}{@{}l@{\hspace{1pt}}*{4}{>{\raggedleft\arraybackslash}X}@{\hspace{6pt}}*{4}{>{\raggedleft\arraybackslash}X}@{\hspace{6pt}}*{3}{>{\columncolor{sysband}[0pt][0pt]\raggedleft\arraybackslash}X}>{\columncolor{sysband}[0pt][2pt]\raggedleft\arraybackslash}X@{\hspace{2pt}}}
    \toprule
    \tup\tdn & \multicolumn{4}{c}{\textbf{HBF-Park}} & \multicolumn{4}{c}{\textbf{NVMe-Park}} & \multicolumn{4}{>{\columncolor{sysband}[0pt][0pt]}c}{\textbf{\sys}} \\
    \cmidrule(lr){2-5}\cmidrule(lr){6-9}\cmidrule(l{2pt}r{2pt}){10-13}
    \tup\tdn\textbf{Sessions/gen.} & 32 & 64 & 128 & 256 & 32 & 64 & 128 & 256 & 32 & 64 & 128 & 256 \\
    \midrule
    \tup Run time (min) & 172 & 279 & 482 & 919 & 174 & 282 & 490 & 940 & \textbf{159} & \textbf{235} & \textbf{316} & \textbf{417} \\
    Wait per call (s) & 11 & 37 & 95 & 203 & 12 & 38 & 97 & 208 & \textbf{6.6} & \textbf{22} & \textbf{47} & \textbf{75} \\
    Median TTFT (s) & 7.8 & 34 & 89 & 187 & 9.3 & 37 & 94 & 188 & \textbf{2.8} & \textbf{22} & \textbf{46} & \textbf{73} \\
    Preemptions & 71 & 170 & 375 & 735 & 63 & 155 & 338 & 706 & \textbf{0} & \textbf{0} & \textbf{0} & \textbf{0} \\
    GB/s per HBF site & \multicolumn{4}{c}{--} & \multicolumn{4}{c}{--} & 123 & 213 & 283 & 305 \\
    TB written & \textbf{1.09} & 2.28 & \textbf{4.70} & \textbf{9.56} & \textbf{1.09} & \textbf{2.27} & \textbf{4.70} & \textbf{9.56} & 1.12 & 2.38 & 4.90 & 9.98 \\
    \tdn Energy (MJ) & 4.1 & 7.3 & 14.0 & 27.4 & 4.2 & 7.6 & 14.6 & 28.7 & \textbf{3.8} & \textbf{6.8} & \textbf{12.3} & \textbf{22.3} \\
    \bottomrule
  \end{tabularx}
\end{table}

\begin{figure}[t]
  \centering
  \includegraphics[width=\columnwidth]{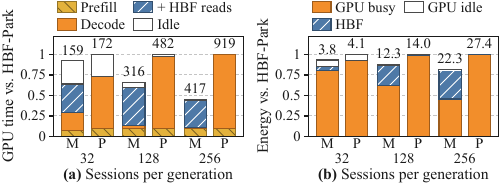}
  \caption{Where (a)~GPU time and (b)~energy go for \sys (M) and HBF-Park (P), each load scaled to HBF-Park's
    total; numbers: totals in min and MJ.}
  \label{fig:breakdown}
\end{figure}

\noindent \textbf{Rising load.} On the DGM loop, reading the ready overflow in place finishes sooner than parking it,
and the gain grows with the load. With HBF used for parking, a ready session whose KV lies beyond HBM waits for room in HBM. A call waits 11--203\,s on average and 71--735 calls are preempted (Table~\ref{tab:gain}). Read in place, the wait falls to 6.6--75\,s and no call is preempted. \sys finishes
8, 16, 34 and 55\% sooner (Fig.~\ref{fig:inplace}); over ten task samples the gain averages 8.0, 17.5, 36.3 and 53.9\%, with 95\% confidence intervals of $\pm$1.5, 1.8, 0.6 and 0.3 points. The saving is
decode time (Fig.~\ref{fig:breakdown}): at 128 sessions \sys decodes for 243\,min, 223 of them with HBF reads, against 421\,min with HBF-Park, and in both designs queueing for admission is 97--100\% of the mean TTFT. \sys spends 0.19 to 9.8\,MJ on HBF to save 0.46 to 14.8\,MJ of GPU energy, uses 9--22\%
less energy than NVMe-Park, and breaks even only if HBF reads cost 33--61\,pJ/bit instead of the 20 we model. Counting HBF standby power, \sys uses less energy than HBF-Park at any value, and less than NVMe-Park up to 55--82\,mW/GB (28--42\,W per stack).
Parking on an NVMe drive instead of HBF changes little (Table~\ref{tab:far}), so \sys finishes 9, 17, 35 and 56\%
sooner than NVMe-Park. The added sites must hold the overflow they serve (Fig.~\ref{fig:inplace}). One HBF
site cannot hold the ready overflow of 128 sessions and takes 2.6\x as long as HBF-Park, so \sys gets one site per 32
sessions. At a thermal limit of 202\,GB/s per site~\cite{hbfsucks}, \sys still finishes 1, 7, 19 and 36\%
sooner than HBF-Park.

\subsection{Ablation Study}
\label{sec:eval-gate}
\label{sec:eval-tail}

\begin{table}[t]
  \centering\footnotesize
  \caption{\sys with one mechanism removed, against \sys (one write per block, WAF 1.00), by sessions per generation. $^a$128 / 256 sessions, SLC sites.
  $^b$No session ends; the LRU variant of \sys (\S\ref{sec:eval-layout}), against itself with them.}
  \label{tab:ablate}
  \definecolor{sysband}{HTML}{E1F5F4}\definecolor{worse}{HTML}{C9302C}%
  \setlength{\tabcolsep}{0pt}\setlength{\aboverulesep}{0pt}\setlength{\belowrulesep}{0pt}%
  \def\tup{\rule{0pt}{2.5ex}}\def\tdn{\rule[-1.0ex]{0pt}{0pt}}\def\w#1{\textcolor{worse}{#1}}%
  \begin{tabularx}{\columnwidth}{@{\hspace{2pt}}l*{4}{>{\raggedleft\arraybackslash}X}@{\hspace{8pt}}l*{4}{>{\raggedleft\arraybackslash}X}@{\hspace{2pt}}}
    \toprule
    \tup\tdn & \multicolumn{4}{c}{\textbf{Run time (\%)}} & \multicolumn{5}{c}{\textbf{Writes and wear}} \\
    \cmidrule(lr){2-5}\cmidrule(l){6-10}
    \tup\tdn\textbf{Sessions/gen.} & 32 & 64 & 128 & 256 & & 32 & 64 & 128 & 256 \\
    \midrule
    \tup No bound (HBF-Greedy)    & \w{+41} & \w{+39} & \w{+29} & \w{+13} & \multicolumn{5}{c}{--} \\
    Never return to HBM        & \w{+61} & \w{+60} & \w{+50} & \w{+28} & \multicolumn{5}{c}{--} \\
    Neither bound nor return   & \w{+55} & \w{+43} & \w{+30} & \w{+12} & \multicolumn{5}{c}{--} \\
    Drop the copy on return    & \multicolumn{4}{c}{\w{+0.5--6}}    & \multicolumn{5}{l}{\w{9--37 writes per block}} \\
    5\,min expiry$^b$          & \multicolumn{4}{c}{\w{+0.2--2.5}}  & \multicolumn{5}{l}{\w{+5--24\% KV written}} \\
    Release only when full$^b$ & \multicolumn{4}{c}{--}         & WAF$^a$ &      &      & \w{2.1}  & \w{2.8}  \\
    No trims, 2\% spare        & \multicolumn{4}{c}{--}         & WAF     & \w{1.11} & \w{1.42} & \w{1.57} & \w{1.77} \\
    \tdn Conservative gate budget   & \multicolumn{4}{c}{within 0.5} & \multicolumn{5}{c}{--} \\
    \bottomrule
  \end{tabularx}
\end{table}

\begin{figure}[t]
  \centering
  \includegraphics[width=\columnwidth]{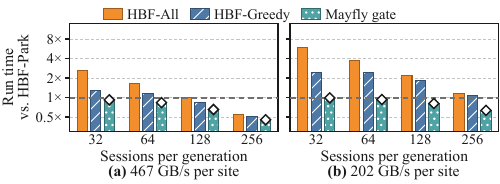}
  \caption{Run time by gate policy over HBF-Park at (a)~467 and (b)~202\,GB/s per site. Diamonds: conservative budget.}
  \label{fig:gate}
\end{figure}

Both parts of the execution path buy run time, and only together (Table~\ref{tab:ablate}). Without the return to HBM, the bound is slower than no bound at all, and each such variant reads 1.5--4.7\x the HBF bytes. The gain is robust to the gate's exact rule (Fig.~\ref{fig:gate}): a conservative budget finishes within 0.5\% of it, while HBF-Greedy takes 1.7--2.6\x as long at 202\,GB/s.
HBF-All takes longer than HBF-Park up to 128 sessions; at 256 it takes 0.55\x as long as HBF-Park, against 0.45\x for \sys.
Against HBF-Greedy, the bound saves 29, 28, 23 and 11\% of the run at 32 to 256 sessions,
and 35, 35, 27 and 11\% over ten task samples: the scarcer the
bandwidth, the more it matters.

\begin{figure}[t]
  \centering
  \includegraphics[width=\columnwidth]{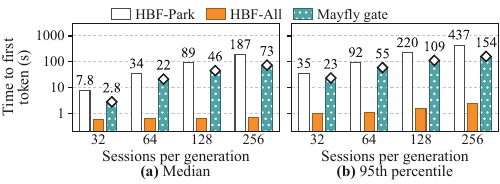}
  \caption{TTFT by gate policy: (a)~median, (b)~95th percentile. Diamonds: conservative budget.}
  \label{fig:gatetail}
\end{figure}

\noindent \textbf{Time to first token.} Against unbounded in-place reads, the bound trades time to first token for run time
(Figs.~\ref{fig:gate}a and~\ref{fig:gatetail}). \sys admits a new call only if its in-place reads do not lower the step's token throughput, and the calls behind it wait in order. At every load it still finishes sooner than HBF-Park and with a
lower 95th-percentile TTFT: 109 against 220\,s at 128 sessions, and 154 against 437\,s at 256. Up to 128 sessions,
designs without a read bound (HBF-All, HBF-Greedy and placing each block once, \S\ref{sec:eval-prior}) keep the 95th-percentile TTFT below 2\,s but finish 1.1--2.8\x later, as reads lengthen each step. At 256, HBF-All has a 95th-percentile TTFT of 2.4\,s and finishes 1.2\x later. Run time is itself a tail measure, as a generation ends with its slowest session; per session, \sys's 99th-percentile duration is 4--54\% below HBF-Park's and 14--30\% below HBF-Greedy's. Against HBF-Park, 97--100\% of sessions finish no later. The gate always admits the oldest running request, so none starves: no running call is held back for longer than 93\,s in total.

\noindent \textbf{Skips.} A running call skips a step when the gate refuses its in-place reads. Of all calls, 14--19\% skip at least one step, which delays the 99th-percentile call by 2.3--2.9\,s over the run, and
under 1\% skip before their first token. A step evaluates the rule 1.6--8.9 times on average, two closed-form estimates each. A gate that misjudges HBF's bandwidth by 20 or 40\% lengthens the run by at most 8 or 21\%. Every such gate finishes sooner than HBF-Park and with a lower 95th-percentile TTFT.

\begin{figure}[t]
  \centering
  \includegraphics[width=\columnwidth]{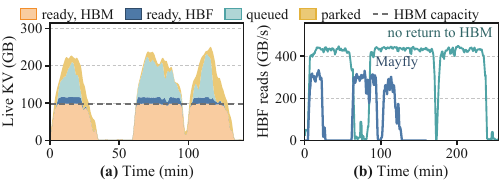}
  \caption{Burst and decline at 32 sessions per generation: (a)~\sys's live KV by state and location; (b)~HBF reads
    with and without the return to HBM.}
  \label{fig:tail}
\end{figure}

\noindent \textbf{Returning KV to HBM as sessions end.} The return confines HBF reads to when live KV
exceeds HBM (Fig.~\ref{fig:tail}a). At 32 sessions per generation, \sys reads HBF in 49--62\% of the GPU's busy time
(Fig.~\ref{fig:tail}b), and in 83 and 92\% at 64 and 128. A variant that never returns KV reads HBF in all of the busy time. It takes 1.28--1.61\x as long and, at 32 sessions, raises the 95th-percentile TTFT from 23 to 82\,s. The time is lost while over a quarter of
a generation's sessions still run, not in its final tail, because sessions end and free HBM throughout the burst. Which return carries the gain depends on the load: at 32 sessions, restoring a parked session as a whole, since returning only KV read in place takes 256 instead of 159\,min; at 64 and 128, keeping KV read in place in HBM, since restoring only parked sessions takes 273 and 447 instead of 235 and 316\,min. The returns themselves cost little: at 32, 128 and 256 sessions they move 24, 49 and 60\,TB from HBF and add 8, 18 and 22\,s, at most 0.13\% of the GPU's busy time.

\subsection{Write Amplification}
\label{sec:eval-writes}
\label{sec:eval-admission}
\label{sec:eval-layout}

A kept copy and trims at session ends hold HBF at one write per block and a WAF of 1.00.
\sys writes a block to HBF when it first leaves HBM and keeps the copy across state switches, and it never writes dead
KV, so each byte written is read 1,040, 2,520 and 4,380 times at 32, 64 and 128 sessions, against 23--28 times when HBF
only parks. Releasing the copy on each return to HBM
costs 0.5--6\% of run time but writes each block 9--37 times. Without end-signal trims, GC copies are
10--44\% of physical writes at 2\% spare with 1\,GB erase units. The WAF grows with load and falls with spare area, up to 1.16, 1.77 and 2.32 at 7, 2 and 1\%; either session or
generation ends bring it to 1.00. On HBF-Park, placement helps only where it follows session lifetimes. A unit per generation reaches 1.00, while streams by KV type~\cite{hbflex} lower the WAF only from 1.60 to 1.54 at 128 sessions and units per
session~\cite{fdp} reach 1.05; 64\,MB units keep one stream below 1.01. On the parking path, trims and
liveness admission cut writes by 15--41\%, while writing a block only when evicted twice~\cite{saria} makes resumed sessions recompute KV, slowing the run by 5--9\%, as Gandalf~\cite{gandalf} finds for history-based gating.

\noindent \textbf{Releasing KV without session ends.} An engine that does not see session ends must release KV by age
or when the tier fills. We measure each such rule on a variant of \sys with LRU tiering in place of its state order and the conservative budget. Given session ends, this variant finishes within 0.5\% of \sys and keeps the WAF at 1.00, so it serves as the reference;
each rule's recompute is charged. No fixed rule matches session ends at
every load. Releasing early loses live KV: expiring after 5\,min also expires sessions waiting on a long tool call, which write 5--24\% more KV and lengthen the run by 0.2--2.5\%; at 256 sessions, expiring blocks instead takes 2.2--3.4\x as long in three of six runs,
as sessions queued for HBM lose their KV and recompute 6--13\x the prefill; and at 1\,min the run takes 3--13\% longer. Releasing late keeps dead KV: expiring after 30\,min,
as SPLASH does~\cite{splash}, keeps half again the KV that the six SLC sites hold at 128 sessions, so the WAF reaches
1.7, and releasing KV only when the tier is full reaches a WAF of 2.1 at 128 sessions and 2.8 at 256. Against the
best expiry for each load in hindsight, untuned session ends come within 1\% of its run time.

\subsection{Sensitivity Analysis}
\label{sec:eval-readpath}
\label{sec:eval-io}
\label{sec:eval-far}
\label{sec:eval-dram}
\label{sec:eval-nvme}

The read path sustains a site's 467\,GB/s unless the array has fewer than 256 planes per die or a page read takes
100\,$\mu$s. We replay the steps that read HBF through array rounds, die-to-base transfer, LDPC and the 9.6\,MB buffer,
layer by layer, with 16 dies per stack. HBFSim~\cite{hbfsim} reports the same bandwidth as our array bound, within 0.2\%, for every
die count, plane count and page read time. With 512 planes of 4\,KiB pages per die, the array delivers 1.68\,TB/s and
adds no step time. With 256 planes it still adds none, while 128 and 64 add 2--7 and 50--83\%, so a thermal limit of 202\,GB/s
corresponds to about 64 planes at full speed. Page reads add nothing up to 50\,$\mu$s and 13--29\% at 100\,$\mu$s. Merging the two paths at 15--33\,$\mu$s per layer adds at most 3.6\%, and a 20 or 100\,$\mu$s stall exposed in every
layer at most 1.0 or 4.3\%. A two-path kernel over host memory runs within
0.993--1.020\x of its slower path on an H200 and matches full attention to $2\times10^{-4}$. Writing HBF at
0.05--0.5\x of its read rate changes run time, energy and drive writes per day by under 1.5\%.

\begin{figure}[t]
  \centering
  \includegraphics[width=\columnwidth]{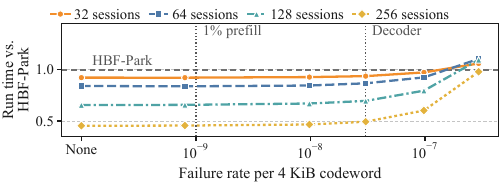}
  \caption{Run time under injected codeword failures.}
  \label{fig:inject}
\end{figure}

\noindent \textbf{Error correction.} A 4\,MiB block spans 1,024 codewords of 4\,KiB, so a per-codeword failure rate of
$10^{-6}$ fails 0.1\% of block reads and would add 3.7--10\x the run's prefill. Holding the added prefill below 1\% takes
about $10^{-9}$. KV is read within 94\,min of its write. At such short retention, a 3D TLC page cycled 2K
times shows up to 19 bit errors per KiB~\cite{park_aspls21}, about $2.3\times10^{-3}$ or 85 per codeword on average,
so every codeword with up to 146 errors must decode. A bit-flipping LDPC decoder for 4\,KiB codewords measures a frame
error rate of $3\times10^{-8}$ at 190 errors with no error floor~\cite{updgdbf}. With failures injected into
every in-place read and the recovery of \S\ref{sec:controller} run (Fig.~\ref{fig:inject}), $8\times10^{-10}$ moves run time by at most 0.5\%, +1\%
takes $3\times10^{-9}$ to $2\times10^{-8}$, and at $3\times10^{-8}$ \sys still finishes 6--51\% sooner than HBF-Park. At 19.3\,Gb/s and 0.12\,mm$^2$ per
decoder in 16\,nm~\cite{updgdbf}, decoding a site's 467\,GB/s takes 23\,mm$^2$, 13--19\% of the base die, and the rated 1.5\,TB/s takes 76\,mm$^2$.

\noindent \textbf{Other far memories.} Off-package memories fall short for two reasons. Host DRAM and an NVMe drive
lack bandwidth: read in place at 128 sessions, they take 22 and 100\x as long as NVMe-Park (Table~\ref{tab:far}). Grace
memory lacks capacity: its 480\,GB cannot hold the ready overflow at 128 sessions, so 71--72\% of the run is recomputed
and it takes 4.1\x as long. At 64 sessions, where it holds the overflow, it trails by 9--10\%. Capacity is a threshold and bandwidth a slope. At a fixed 450\,GB/s, 480\,GB takes 4.29\x as long as \sys and a terabyte
suffices. At a fixed 2\,TB, 450\,GB/s takes up to 1.38\x as long.

\subsection{Endurance Analysis}
\label{sec:eval-endurance}

A stack's life follows its write control (\S\ref{sec:eval-writes}). With write control, a TLC stack absorbs 20, 14, 11 and 8.4 drive writes per day at 32 to 256 sessions with a WAF
of 1.00, so each block spends that many of its 7.5K P/E cycles~\cite{star} a day. The kept copy makes a stack last
9--37\x as many generations as releasing it at each return to HBM. Session ends make it last 1.7--2.8\x as
long as a 30-min expiry or release when the tier is full. KV is read young, at most
94\,min old, and relaxing the unused retention~\cite{warm,retention_relax} buys 2.0--3.8\x more cycles by
Luo et al.'s model~\cite{luo3d}; our wear figures exclude it.

SLC mode trades two thirds of the capacity for endurance, and more sites add bandwidth: 4, 4, 6 and 12 SLC sites finish 9--13\% sooner than TLC at 32 to 256 sessions, and hold parked and
ready KV alike, with no drive below. They absorb 17--24 drive writes per day, so a generation uses 0.6--1.7 of a
block's P/E cycles, against 30K cycles for 3D TLC run as pSLC at a year of end-of-life
retention~\cite{swissbit_pslc}, 100K with no retention condition stated~\cite{innodisk_3ie7}, and over 30K for
native 3D SLC~\cite{kioxia_fl6}; at 30K cycles a pSLC stack lasts 3.4--4.8 years under continuous load. Wear-out loses
no data: a worn block leaves the tier, and the KV it held is recomputed from its context, as after a failed read. The conservative operating point takes pSLC sites at 202\,GB/s, which an array of 128 planes per die
sustains, with 30K cycles: 4, 4, 6 and 12 sites finish 1.12, 1.17, 1.39 and 1.75\x faster than NVMe-Park at
32 to 256 sessions, with a 1.5--6.8\x lower median TTFT, 9--15\% less energy
and no preemptions.

\subsection{Generality}
\label{sec:eval-agents}
\label{sec:eval-threshold}

\begin{figure}[t]
  \centering
  \includegraphics[width=\columnwidth]{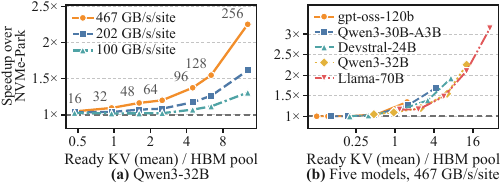}
  \caption{Gain over NVMe-Park by how far ready KV outgrows HBM: (a)~Qwen3-32B by site bandwidth (labels:
    sessions); (b)~five models at 467\,GB/s, ordered by KV per token.}
  \label{fig:threshold}
\end{figure}

The gain tracks not the number of sessions but how far ready KV outgrows HBM (Fig.~\ref{fig:threshold}). At
467\,GB/s per site, \sys runs 1.05\x faster than NVMe-Park while ready KV averages 0.47\x the HBM pool, 1.20\x
at 2.5\x the pool, 1.55\x at 6.4\x and 2.25\x at 12.8\x. At 100\,GB/s the gain stays under 10\% until ready KV averages
about 5\x the pool. The model sets which load reaches that point: gpt-oss-120b~\cite{gptoss}, a mixture-of-experts model, keeps 36\,KiB of KV per
token, so its ready KV fits in HBM up to 64 sessions, where \sys and every baseline tie, and \sys is 1.05\x faster than
NVMe-Park at 128 and 1.35\x at 256. Llama-3.3-70B keeps 320\,KiB, so \sys is 1.16\x faster at 16 sessions and 3.14\x at 256. Against
the same ratio, the five models rise from parity below a fifth of the pool to 1.10--1.35\x near it and 1.39--3.14\x
beyond three times it (Fig.~\ref{fig:threshold}b). The strongest baseline is HBF-Park at low load and HBF-Greedy elsewhere. Against it, \sys finishes 1.08--1.29\x faster for Qwen3-32B, 1.12--1.41\x for
Llama-3.3-70B, 1.12--1.14\x for Devstral-24B and 1.03\x for Qwen3-30B-A3B~\cite{qwen3} up to 128 sessions, and
1.00--1.13\x at 256, whose eight sites together read faster than HBM. While ready KV averages under half the pool,
HBF-Greedy finishes within 0.4\% of \sys. From about the pool on, it takes 3--71\% longer up to 128 sessions. FP8 KV halves the bytes per token and composes with \sys, which then finishes 1.32\x faster
than the best FP8 baseline at 128 sessions (three-sample means); we keep BF16 KV, as FP8 KV changed 41 of 530 MATH
answers and lowered LongBench from 43.4 to 39.6 in our runs.
As concurrent autonomous agents, with $8V$ busy slots and no generations, the sessions finish 9, 22, 39 and
56\% sooner with \sys than with NVMe-Park at 32 to 256 slots (Fig.~\ref{fig:real}), as they keep ready KV beyond HBM
throughout and RSI only in bursts. For these agents a fixed expiry suffices: with 5 or 30\,min they
finish within 1\% of \sys at 128 slots and write 1--6\% more to HBF.
In RSI, no fixed expiry fits every load (\S\ref{sec:eval-layout}), so \sys relies on
the session ends that the orchestrator announces.

\subsection{Comparison with Prior Work}
\label{sec:eval-prior}
\label{sec:eval-dropped}
\label{sec:eval-robust}

With Qwen3-32B and one site per 32 sessions, every prior HBF tier takes longer than \sys at 32 to 256 sessions.

\noindent \textbf{Prior HBF tiers.} We run adapted versions of the closest HBF tiers' policies
(\S\ref{sec:bg-offload}) on the same sites, traces and step model; tiers outside Tables~\ref{tab:gain} and~\ref{tab:ablate} are means over three task samples. HBF-Park follows Baek~et~al.~\cite{baek_hbf} and takes 1.08, 1.19, 1.52 and 2.20\x as long as \sys at 32 to 256 sessions. A
FlashAccel-style tier~\cite{flashaccel} writes every evicted block through, reads all ready KV in place and frees the
oldest sessions' KV when it fills. Without its copies of hot pages into HBM it takes 1.20--2.90\x \sys's time, and with
HBM filled at no cost, which is HBF-Greedy, as Yu~et~al.~\cite{yu_hbf} without prefix reloads, 13--41\% longer (Table~\ref{tab:ablate}). A Lachesis-style placement~\cite{lachesis} fixes each block's tier at
allocation and frees it when its session ends. It takes 1.13--1.57\x as long with HBM first and 1.16--2.08\x with
oracle lifetimes. Without copies, both keep reading the same blocks from HBF, at 374--448 of a site's 467\,GB/s, and use 1.6--3.9\x \sys's energy up to 128 sessions. SETA~\cite{seta}, which serves sessions past 32K tokens from HBF and recalls them to HBM under
load, and HBFlex~\cite{hbflex}, which reads all ready KV from HBF, take 1.33 and 1.59\x as long as \sys at 128 sessions on four HBF sites.

\noindent \textbf{Concurrency limits.} Capping concurrency does not rescue HBM alone: at its best cap,
eight sessions, it takes 1.20--3.19\x \sys's time at 32 to 256 sessions. At 128 sessions on four sites, capping HBF-Park at 32 to 128 sessions changes its run time by 3\%, as a call that waited for room in HBM now
waits for a slot. At a matched 95th-percentile TTFT \sys finishes 1.27--1.64\x faster than either parking design.
Capped at 32 sessions, its best cap, HBF-Greedy takes 363\,min; uncapped, \sys is 12\% faster with no cap to tune.

\section{Conclusion}
\label{sec:conclusion}

Harness RSI's KV overflows HBM in bursts, at high load mostly as ready sessions' KV. Parked KV needs only
capacity, which any tier provides; ready KV beyond HBM needs execution-time bandwidth, which HBF provides on the GPU
package. \sys keeps one HBF copy valid across state switches, bounds HBF reads per
step with a return to HBM as sessions end, and invalidates KV at announced session ends. On a simulated H200, it finishes
34\% sooner than parking on HBF at 128 sessions (55\% at 256 on eight sites) and 23\% sooner than reading HBF whenever HBM is full. The
mechanisms apply wherever ready KV exceeds HBM, as it does when autonomous agents run concurrently.

\bibliographystyle{ACM-Reference-Format}
\bibliography{refs}

@inproceedings{dgm,
  author = {Jenny Zhang and Shengran Hu and Cong Lu and Robert Tjarko Lange and Jeff Clune},
  title = {Darwin {G}{\"o}del Machine: Open-Ended Evolution of Self-Improving Agents},
  booktitle = {International Conference on Learning Representations (ICLR)},
  year = {2026}
}

@inproceedings{hgm,
  author = {Wenyi Wang and Piotr Pi{\k{e}}kos and Nanbo Li and Firas Laakom and Yimeng Chen and Mateusz Ostaszewski and Mingchen Zhuge and J{\"u}rgen Schmidhuber},
  title = {{Huxley-G{\"o}del} Machine: Human-Level Coding Agent Development by an Approximation of the Optimal Self-Improving Machine},
  booktitle = {International Conference on Learning Representations (ICLR)},
  year = {2026}
}

@inproceedings{adas,
  author = {Shengran Hu and Cong Lu and Jeff Clune},
  title = {Automated Design of Agentic Systems},
  booktitle = {International Conference on Learning Representations (ICLR)},
  year = {2025}
}

@inproceedings{godelagent,
  author = {Xunjian Yin and Xinyi Wang and Liangming Pan and Li Lin and Xiaojun Wan and William Yang Wang},
  title = {{G{\"o}del} Agent: A Self-Referential Agent Framework for Recursively Self-Improvement},
  booktitle = {Proceedings of the 63rd Annual Meeting of the Association for Computational Linguistics (ACL)},
  pages = {27890--27913},
  year = {2025},
  doi = {10.18653/v1/2025.acl-long.1354}
}

@inproceedings{stop,
  author = {Eric Zelikman and Eliana Lorch and Lester Mackey and Adam Tauman Kalai},
  title = {Self-Taught Optimizer ({STOP}): Recursively Self-Improving Code Generation},
  booktitle = {Conference on Language Modeling (COLM)},
  year = {2024}
}

@article{alphaevolve,
  author = {Alexander Novikov and Ng{\^a}n V{\~u} and Marvin Eisenberger and Emilien Dupont and Po-Sen Huang and Adam Zsolt Wagner and Sergey Shirobokov and Borislav Kozlovskii and Francisco J. R. Ruiz and Abbas Mehrabian and M. Pawan Kumar and Abigail See and Swarat Chaudhuri and George Holland and Alex Davies and Sebastian Nowozin and Pushmeet Kohli and Matej Balog},
  title = {{AlphaEvolve}: A Coding Agent for Scientific and Algorithmic Discovery},
  journal = {arXiv preprint arXiv:2506.13131},
  year = {2025}
}

@article{rzero,
  author = {Chengsong Huang and Wenhao Yu and Xiaoyang Wang and Hongming Zhang and Zongxia Li and Ruosen Li and Jiaxin Huang and Haitao Mi and Dong Yu},
  title = {{R-Zero}: Self-Evolving Reasoning {LLM} from Zero Data},
  journal = {arXiv preprint arXiv:2508.05004},
  year = {2025}
}

@misc{openevolve,
  author = {Asankhaya Sharma},
  title = {{OpenEvolve}: An Open-Source Evolutionary Coding Agent},
  howpublished = {\url{https://github.com/algorithmicsuperintelligence/openevolve}},
  publisher = {GitHub},
  year = {2025}
}

@inproceedings{swebench,
  author = {Carlos E. Jimenez and John Yang and Alexander Wettig and Shunyu Yao and Kexin Pei and Ofir Press and Karthik Narasimhan},
  title = {{SWE-bench}: Can Language Models Resolve Real-World {GitHub} Issues?},
  booktitle = {International Conference on Learning Representations (ICLR)},
  year = {2024}
}

@inproceedings{swerebench,
  author = {Ibragim Badertdinov and Alexander Golubev and Maksim Nekrashevich and Anton Shevtsov and Simon Karasik and Andrei Andriushchenko and Maria Trofimova and Daria Litvintseva and Boris Yangel},
  title = {{SWE-rebench}: An Automated Pipeline for Task Collection and Decontaminated Evaluation of Software Engineering Agents},
  booktitle = {Advances in Neural Information Processing Systems (NeurIPS), Datasets and Benchmarks Track},
  year = {2025},
  note = {Trajectories: nebius/SWE-rebench-openhands-trajectories}
}

@inproceedings{openhands,
  author = {Xingyao Wang and Boxuan Li and Yufan Song and Frank F. Xu and Xiangru Tang and Mingchen Zhuge and Jiayi Pan and Yueqi Song and Bowen Li and Jaskirat Singh and Hoang H. Tran and Fuqiang Li and Ren Ma and Mingzhang Zheng and Bill Qian and Yanjun Shao and Niklas Muennighoff and Yizhe Zhang and Binyuan Hui and Junyang Lin and Robert Brennan and Hao Peng and Heng Ji and Graham Neubig},
  title = {{OpenHands}: An Open Platform for {AI} Software Developers as Generalist Agents},
  booktitle = {International Conference on Learning Representations (ICLR)},
  year = {2025}
}

@article{tracelab,
  author = {Kan Zhu and Mathew Jacob and Chenxi Ma and Yi Pan and Stephanie Wang and Arvind Krishnamurthy and Baris Kasikci},
  title = {{TraceLab}: Characterizing Coding Agent Workloads for {LLM} Serving},
  journal = {arXiv preprint arXiv:2606.30560},
  year = {2026},
  note = {SyFI coding trace (syfi\_coding\_trace), CC BY 4.0, \url{https://github.com/uw-syfi/TraceLab}}
}

@inproceedings{vllm,
  author = {Woosuk Kwon and Zhuohan Li and Siyuan Zhuang and Ying Sheng and Lianmin Zheng and Cody Hao Yu and Joseph E. Gonzalez and Hao Zhang and Ion Stoica},
  title = {Efficient Memory Management for Large Language Model Serving with {PagedAttention}},
  booktitle = {ACM Symposium on Operating Systems Principles (SOSP)},
  year = {2023}
}

@inproceedings{sglang,
  author = {Lianmin Zheng and Liangsheng Yin and Zhiqiang Xie and Chuyue Sun and Jeff Huang and Cody Hao Yu and Shiyi Cao and Christos Kozyrakis and Ion Stoica and Joseph E. Gonzalez and Clark Barrett and Ying Sheng},
  title = {{SGLang}: Efficient Execution of Structured Language Model Programs},
  booktitle = {Advances in Neural Information Processing Systems (NeurIPS)},
  year = {2024}
}

@inproceedings{infercept,
  author = {Reyna Abhyankar and Zijian He and Vikranth Srivatsa and Hao Zhang and Yiying Zhang},
  title = {{InferCept}: Efficient Intercept Support for Augmented Large Language Model Inference},
  booktitle = {International Conference on Machine Learning (ICML)},
  series = {PMLR},
  volume = {235},
  pages = {81--95},
  year = {2024}
}

@inproceedings{kvflow,
  author = {Zaifeng Pan and Ajjkumar Dahyalal Patel and Yipeng Shen and Zhengding Hu and Yue Guan and Wan-Lu Li and Lianhui Qin and Yida Wang and Yufei Ding},
  title = {{KVFlow}: Efficient Prefix Caching for Accelerating {LLM}-Based Multi-Agent Workflows},
  booktitle = {Advances in Neural Information Processing Systems (NeurIPS)},
  year = {2025}
}

@article{continuum,
  author = {Hanchen Li and Runyuan He and Qiuyang Mang and Qizheng Zhang and Huanzhi Mao and Xiaokun Chen and Hangrui Zhou and Huanchen Zhang and Alvin Cheung and Joseph Gonzalez and Ion Stoica},
  title = {Continuum: Efficient and Robust Multi-Turn {LLM} Agent Scheduling with {KV} Cache Time-to-Live},
  journal = {arXiv preprint arXiv:2511.02230},
  year = {2025}
}

@article{lmcache,
  author = {Yuhan Liu and Yihua Cheng and Jiayi Yao and Yuwei An and Xiaokun Chen and Shaoting Feng and Yuyang Huang and Samuel Shen and Rui Zhang and Kuntai Du and Junchen Jiang},
  title = {{LMCache}: An Efficient {KV} Cache Layer for Enterprise-Scale {LLM} Inference},
  journal = {arXiv preprint arXiv:2510.09665},
  year = {2025}
}

@inproceedings{cachedattention,
  author = {Bin Gao and Zhuomin He and Puru Sharma and Qingxuan Kang and Djordje Jevdjic and Junbo Deng and Xingkun Yang and Zhou Yu and Pengfei Zuo},
  title = {Cost-Efficient Large Language Model Serving for Multi-turn Conversations with {CachedAttention}},
  booktitle = {USENIX Annual Technical Conference (ATC)},
  year = {2024}
}

@inproceedings{mooncake,
  author = {Ruoyu Qin and Zheming Li and Weiran He and Jialei Cui and Feng Ren and Mingxing Zhang and Yongwei Wu and Weimin Zheng and Xinran Xu},
  title = {Mooncake: Trading More Storage for Less Computation --- A {KVCache}-centric Architecture for Serving {LLM} Chatbot},
  booktitle = {USENIX Conference on File and Storage Technologies (FAST)},
  year = {2025}
}

@misc{ocphbf,
  author = {{SK hynix} and {Sandisk}},
  title = {{OCP} High Bandwidth Flash ({HBF}) Architecture Specification, Version 0.7.0},
  howpublished = {Open Compute Project (OCP)},
  month = aug,
  year = {2026},
  url = {https://www.opencompute.org/documents/ocp-hbf-architecture-specification-v0-7-0-final-pdf/},
  note = {Released at FMS 2026}
}

@article{patterson_hbf,
  author = {Xiaoyu Ma and David Patterson},
  title = {Challenges and Research Directions for Large Language Model Inference Hardware},
  journal = {Computer},
  volume = {59},
  number = {5},
  pages = {55--64},
  year = {2026},
  doi = {10.1109/MC.2026.3652916},
  note = {arXiv:2601.05047}
}

@article{hbfsucks,
  author = {Zhuoran Li and Zhuohang Bian and Xin Huang and Yibo Zhao and Guangyu Sun and Youwei Zhuo},
  title = {{HBF} Sucks? {A} Full-Stack Characterization of High-Bandwidth Flash for {KV}-Centric {LLM} Serving},
  journal = {arXiv preprint arXiv:2608.11668},
  year = {2026}
}

@article{hbfsim,
  author = {Yanpeng Hu and Yiwei Yang and Yuanwu Zhu and Yusheng Zheng and Andi Quinn and Wei Zhang},
  title = {{HBFSim}: Fast and Faithful Simulation of High-Bandwidth Flash Under Real {GPU} Execution},
  journal = {arXiv preprint arXiv:2609.09800},
  year = {2026}
}

@article{flashaccel,
  author = {Xinyu Wang and Yalong Xue and Xiaotian Sun and Xiaoyu Zhang and Xinjiang Zhang and Chunmeng Dou and Xueqi Li and Xiaoming Chen},
  title = {{FlashAccel}: Leveraging High-Bandwidth Flash ({HBF}) for High-Throughput {LLM} Inference},
  journal = {arXiv preprint arXiv:2607.10186},
  year = {2026}
}

@article{baek_hbf,
  author = {Jongjin Baek and Won Ji and Seungjae Yoo and Joo-Young Kim},
  title = {Hot-Cold Tiering of {HBM} and High Bandwidth Flash for Agentic {LLM} Serving},
  journal = {IEEE Computer Architecture Letters},
  volume = {25},
  number = {2},
  pages = {355--358},
  year = {2026},
  doi = {10.1109/LCA.2026.3729099},
  note = {arXiv:2609.25782}
}

@article{yu_hbf,
  author = {Zack Yu and Chloe Wong and Coleman Hooper and Minjae Lee and Wonjun Kang and Youngjin Cho and Michael W. Mahoney and Yakun Sophia Shao and Kurt Keutzer and Amir Gholami},
  title = {Characterizing High Bandwidth Flash for {LLM} Serving},
  journal = {arXiv preprint arXiv:2609.39131},
  year = {2026}
}

@article{hbflex,
  author = {Shuzhang Zhong and Weikai Xu and Yifan Zhou and Tongbin Zhao and Tenghao Zhao and Yifei Kang and Cunyin Chang and Shu Li and Guangyu Sun and Meng Li},
  title = {{HBFlex}: A Flexible Memory System for Bridging Fine-Grained {LLM} States and Coarse-Grained {HBF} Parallel Execution},
  journal = {arXiv preprint arXiv:2609.18675},
  year = {2026}
}

@article{lachesis,
  author = {Jaehoon Yang and Jeongmin Lee and Haneul Park and Seung Yul Lee and Nam Sung Kim and Jae W. Lee},
  title = {Lachesis: Lifetime-Aware {KV} Cache Placement for Agent Serving across {HBM} and High-Bandwidth Flash},
  journal = {arXiv preprint arXiv:2610.08378},
  year = {2026}
}

@article{splash,
  author = {Aditya Anirudh Jonnalagadda and Agasthi Haputhanthri and Pranav Dangi and Rohan Juneja and Wenshuo Yue and Aritra Bagchi and Bin Gao and Tulika Mitra},
  title = {{SPLASH}: Co-Designing Sparse Attention with High-Bandwidth Flash for Efficient Long-Context Inference},
  journal = {arXiv preprint arXiv:2609.23816},
  year = {2026}
}

@article{peng_hbf,
  author = {Danni Peng and Kai Wu and Tianyu Zuo and Pengfei Xia and Hui Zang},
  title = {Enabling High-Bandwidth Flash for Generative Recommendation Serving with Write-Aware {KV} Cache Policy},
  journal = {arXiv preprint arXiv:2609.07175},
  year = {2026}
}

@article{memexplorer,
  author = {Haoran Wu and Zeyu Cao and Yao Lai and Binglei Lou and Jiayi Nie and Can Xiao and Timi Adeniran and Kevin Lau and Przemyslaw Forys and Kauser Johar and Catriona Wright and Junyi Liu and Kai Shi and Nicholas D. Lane and Rika Antonova and Jianyi Cheng and Timothy Jones and Aaron Zhao and Robert Mullins},
  title = {{MemExplorer}: Navigating the Heterogeneous Memory Design Space for Agentic Inference {NPUs}},
  journal = {arXiv preprint arXiv:2604.16007},
  year = {2026}
}

@article{gandalf,
  author = {Maharani A. P. Irawan and Kuntai Du and Jiayi Yao and Dongjoo Seo and Yuyang Huang and Haryadi S. Gunawi and Junchen Jiang},
  title = {Gandalf: Flash Shall Not Die! {I}ncreasing Flash Storage Lifetime for {KV} Cache Offloading in {LLM} Serving},
  journal = {arXiv preprint arXiv:2610.30791},
  year = {2026}
}

@article{saria,
  author = {Hui Li and Qihui Zhou and Xiao Yan and Shumeng Zhang and Zijie Dai and Zequn Gong and Mingcong Song and Guoping Long and Hongjie Si and James Cheng},
  title = {{SARIA}: Persistence-Aware {KV} Caching with Coordinated {DRAM}--{SSD} Reload for {LLM} Agents},
  journal = {arXiv preprint arXiv:2610.31169},
  year = {2026}
}

@article{qwen3,
  author = {An Yang and Anfeng Li and Baosong Yang and Beichen Zhang and Binyuan Hui and Bo Zheng and Bowen Yu and Chang Gao and Chengen Huang and Chenxu Lv and Chujie Zheng and Dayiheng Liu and Fan Zhou and Fei Huang and Feng Hu and Hao Ge and Haoran Wei and Huan Lin and Jialong Tang and Jian Yang and Jianhong Tu and Jianwei Zhang and Jianxin Yang and Jiaxi Yang and Jing Zhou and Jingren Zhou and Junyang Lin and Kai Dang and Keqin Bao and Kexin Yang and Le Yu and Lianghao Deng and Mei Li and Mingfeng Xue and Mingze Li and Pei Zhang and Peng Wang and Qin Zhu and Rui Men and Ruize Gao and Shixuan Liu and Shuang Luo and Tianhao Li and Tianyi Tang and Wenbiao Yin and Xingzhang Ren and Xinyu Wang and Xinyu Zhang and Xuancheng Ren and Yang Fan and Yang Su and Yichang Zhang and Yinger Zhang and Yu Wan and Yuqiong Liu and Zekun Wang and Zeyu Cui and Zhenru Zhang and Zhipeng Zhou and Zihan Qiu},
  title = {Qwen3 Technical Report},
  journal = {arXiv preprint arXiv:2505.09388},
  year = {2025}
}

@article{gptoss,
  author = {{OpenAI}},
  title = {gpt-oss-120b \& gpt-oss-20b Model Card},
  journal = {arXiv preprint arXiv:2508.10925},
  year = {2025}
}

@inproceedings{warm,
  author = {Yixin Luo and Yu Cai and Saugata Ghose and Jongmoo Choi and Onur Mutlu},
  title = {{WARM}: Improving {NAND} Flash Memory Lifetime with Write-Hotness Aware Retention Management},
  booktitle = {IEEE Conference on Mass Storage Systems and Technologies (MSST)},
  year = {2015},
  doi = {10.1109/MSST.2015.7208284}
}

@inproceedings{retention_relax,
  author = {Ren-Shuo Liu and Chia-Lin Yang and Wei Wu},
  title = {Optimizing {NAND} Flash-Based {SSDs} via Retention Relaxation},
  booktitle = {USENIX Conference on File and Storage Technologies (FAST)},
  year = {2012}
}

@inproceedings{luo3d,
  author = {Yixin Luo and Saugata Ghose and Yu Cai and Erich F. Haratsch and Onur Mutlu},
  title = {Improving {3D} {NAND} Flash Memory Lifetime by Tolerating Early Retention Loss and Process Variation},
  booktitle = {ACM SIGMETRICS},
  year = {2018}
}

@misc{fdp,
  author = {{NVM Express}},
  title = {{TP4146} Flexible Data Placement},
  howpublished = {NVM Express ratified technical proposal; incorporated in NVM Express Base Specification 2.1},
  year = {2022}
}

@article{son_hbf,
  author = {Dowon Son and Yonggon Park and Hyunuk Cho and Hyungkyu Ham and Onur Mutlu and Sungjin Lee and Gwangsun Kim and Jisung Park},
  title = {Exploring High-Bandwidth Flash for Modern {LLM} Inference: Opportunities and Challenges},
  journal = {IEEE Computer Architecture Letters},
  year = {2026},
  doi = {10.1109/LCA.2026.3705817},
  note = {arXiv:2608.13868}
}

@inproceedings{park_aspls21,
  author = {Jisung Park and Myungsuk Kim and Myoungjun Chun and Lois Orosa and Jihong Kim and Onur Mutlu},
  title = {Reducing Solid-State Drive Read Latency by Optimizing Read-Retry},
  booktitle = {Proceedings of the 26th ACM International Conference on Architectural Support for Programming Languages and Operating Systems (ASPLOS)},
  year = {2021}
}

@article{updgdbf,
  author = {Li-Wei Liu and Yen-Chin Liao and Hsie-Chia Chang},
  title = {{UP-GDBF}: A 19.3 Gbps Error Floor Free 4{KB} {LDPC} Decoder for {NAND} Flash Applications},
  journal = {IEEE Open Journal of Circuits and Systems},
  volume = {3},
  pages = {228--236},
  year = {2022},
  doi = {10.1109/OJCAS.2022.3209152}
}

@inproceedings{star,
  author = {Omin Kwon and Kyungjun Oh and Jaeyong Lee and Myungsuk Kim and Jihong Kim},
  title = {{STAR}: Improving Lifetime and Performance of High-Capacity Modern {SSDs} Using State-Aware Randomizer},
  booktitle = {Proceedings of the IEEE/ACM International Conference on Computer-Aided Design (ICCAD)},
  pages = {1--9},
  year = {2025}
}

@inproceedings{aero,
  author = {Sungjun Cho and Beomjun Kim and Hyunuk Cho and Gyeongseob Seo and Onur Mutlu and Myungsuk Kim and Jisung Park},
  title = {{AERO}: Adaptive Erase Operation for Improving Lifetime and Performance of Modern {NAND} Flash-Based {SSDs}},
  booktitle = {Proceedings of the 29th ACM International Conference on Architectural Support for Programming Languages and Operating Systems (ASPLOS)},
  year = {2024}
}

@article{hypaic,
  author = {Minho Ha and Myunghyun Rhee and Euiseok Kim and Hoshik Kim},
  title = {{HyPAIC}: Personal {AI} Computer with High Bandwidth Flash for High Performance Trillion-Parameter {LLM} Inference},
  journal = {arXiv preprint arXiv:2611.04144},
  year = {2026}
}

@misc{flashdecoding,
  author = {Tri Dao and Daniel Haziza and Francisco Massa and Grigory Sizov},
  title = {{Flash-Decoding} for Long-Context Inference},
  howpublished = {\url{https://pytorch.org/blog/flash-decoding/}},
  month = oct,
  year = {2023}
}

@misc{innodisk_3ie7,
  author = {{Innodisk}},
  title = {{M.2 (S80) 3IE7} Industrial {SSD} Datasheet},
  note = {iSLC on 112-layer 3D TLC NAND, 100K program/erase cycles},
  howpublished = {\url{https://www.innodisk.com/upload/file/Innodisk_M2_S80_3IE7_Datasheet.pdf}}
}

@misc{kioxia_fl6,
  author = {{KIOXIA}},
  title = {{FL6} Series Enterprise {NVMe} {SSD} Product Brief},
  note = {XL-FLASH, 96-layer BiCS 3D SLC; 60 DWPD over five years},
  howpublished = {\url{https://americas.kioxia.com/content/dam/kioxia/shared/business/ssd/enterprise-ssd/asset/productbrief/eSSD-FL6-product-brief.pdf}},
  year = {2023}
}

@misc{swissbit_pslc,
  author = {{Swissbit}},
  title = {{X-76m2} Industrial {M.2} {SSD} Data Sheet},
  note = {3D NAND in pseudo-SLC mode, 30K program/erase cycles},
  year = {2023}
}

@misc{harness2,
  author = {Ruiyao Xu and Yanfei Chen and Zhongying CuiZhu and Bhavana Dalvi Mishra and Yifei Ming and Han Yu and Rujun Han and Chen-Yu Lee and Tomas Pfister},
  title = {{Harness$^2$}: Recursive Agent Harnessing for an Open World},
  howpublished = {arXiv:2610.14224},
  year = {2026}
}

@incollection{good1965,
  author = {Irving John Good},
  title = {Speculations Concerning the First Ultraintelligent Machine},
  booktitle = {Advances in Computers},
  editor = {Franz L. Alt and Morris Rubinoff},
  volume = {6},
  pages = {31--88},
  publisher = {Elsevier},
  year = {1966}
}

@incollection{godelmachine,
  author = {J{\"u}rgen Schmidhuber},
  title = {{G}{\"o}del Machines: Fully Self-Referential Optimal Universal Self-Improvers},
  booktitle = {Artificial General Intelligence},
  editor = {Ben Goertzel and Cassio Pennachin},
  series = {Cognitive Technologies},
  publisher = {Springer},
  pages = {199--226},
  year = {2007}
}

@inproceedings{promptbreeder,
  author = {Chrisantha Fernando and Dylan Banarse and Henryk Michalewski and Simon Osindero and Tim Rockt{\"a}schel},
  title = {Promptbreeder: Self-Referential Self-Improvement via Prompt Evolution},
  booktitle = {International Conference on Machine Learning (ICML)},
  year = {2024}
}

@article{funsearch,
  author = {Bernardino Romera-Paredes and Mohammadamin Barekatain and Alexander Novikov and Matej Balog and M. Pawan Kumar and Emilien Dupont and Francisco J. R. Ruiz and Jordan S. Ellenberg and Pengming Wang and Omar Fawzi and Pushmeet Kohli and Alhussein Fawzi},
  title = {Mathematical Discoveries from Program Search with Large Language Models},
  journal = {Nature},
  volume = {625},
  number = {7995},
  pages = {468--475},
  year = {2024}
}

@article{metaharness,
  author = {Yoonho Lee and Roshen Nair and Qizheng Zhang and Kangwook Lee and Omar Khattab and Chelsea Finn},
  title = {{Meta-Harness}: End-to-End Optimization of Model Harnesses},
  journal = {arXiv preprint arXiv:2603.28052},
  year = {2026}
}

@article{ahe,
  author = {Jiahang Lin and Shichun Liu and Chengjun Pan and Lizhi Lin and Shihan Dou and Zhiheng Xi and Xuanjing Huang and Hang Yan and Zhenhua Han and Tao Gui and Yu-Gang Jiang},
  title = {Agentic Harness Engineering: Observability-Driven Automatic Evolution of Coding-Agent Harnesses},
  journal = {arXiv preprint arXiv:2604.25850},
  year = {2026}
}

@article{sift,
  author = {Xinghong Fu and Aravinth Kulanthaivelu and Yutaro Yamada},
  title = {Self Improvement via Fast Tree-Search},
  journal = {arXiv preprint arXiv:2609.19526},
  year = {2026}
}

@article{rrsi,
  author = {Peng Xia and Rujun Han and Zifeng Wang and Yanfei Chen and Yufan Zhuang and Yoonho Lee and Chengsong Huang and Han Yu and Zhongying CuiZhu and Yifei Ming and Huaxiu Yao and Burak Gokturk and Tomas Pfister and Chen-Yu Lee},
  title = {{RRSI}: Regularized Recursive Self-Improvement of Agent Harnesses},
  journal = {arXiv preprint arXiv:2609.24972},
  year = {2026}
}

@inproceedings{neo,
  author = {Xuanlin Jiang and Yang Zhou and Shiyi Cao and Ion Stoica and Minlan Yu},
  title = {{NEO}: Saving {GPU} Memory Crisis with {CPU} Offloading for Online {LLM} Inference},
  booktitle = {Conference on Machine Learning and Systems (MLSys)},
  year = {2025}
}

@inproceedings{impress,
  author = {Weijian Chen and Shuibing He and Haoyang Qu and Ruidong Zhang and Siling Yang and Ping Chen and Yi Zheng and Baoxing Huai and Gang Chen},
  title = {{IMPRESS}: An Importance-Informed Multi-Tier Prefix {KV} Storage System for Large Language Model Inference},
  booktitle = {USENIX Conference on File and Storage Technologies (FAST)},
  year = {2025}
}

@article{instinfer,
  author = {Xiurui Pan and Endian Li and Qiao Li and Shengwen Liang and Yizhou Shan and Ke Zhou and Yingwei Luo and Xiaolin Wang and Jie Zhang},
  title = {{InstInfer}: In-Storage Attention Offloading for Cost-Effective Long-Context {LLM} Inference},
  journal = {arXiv preprint arXiv:2409.04992},
  year = {2024}
}

@inproceedings{hilos,
  author = {Hongsun Jang and Jaeyong Song and Changmin Shin and Si Ung Noh and Jaewon Jung and Jisung Park and Jinho Lee},
  title = {A Cost-Effective Near-Storage Processing Solution for Offline Inference of Long-Context {LLMs}},
  booktitle = {ACM International Conference on Architectural Support for Programming Languages and Operating Systems (ASPLOS)},
  year = {2026}
}

@article{tutti,
  author = {Shi Qiu and Yifan Hu and Xintao Wang and Wenhao Zhu and Jianqin Yan and Hao Chen and Kaiqiang Xu and Kai Chen and Yiming Zhang},
  title = {Tutti: Making {SSD}-Backed {KV} Cache Practical for Long-Context {LLM} Serving},
  journal = {arXiv preprint arXiv:2605.03375},
  year = {2026}
}

@article{pykvcache,
  author = {Joseph Kanichai and Tiziano De Matteis and Animesh Trivedi},
  title = {Building py-kvcache: A Performance Characterization of External {KV} Caching for {vLLM} with {NVMe} {SSDs}},
  journal = {arXiv preprint arXiv:2609.11744},
  year = {2026}
}

@article{seta,
  author = {Haneul Park and Jiwon Chang and Wonkyoo Lee and Dongsop Lee and Chunsung Kim and Nam Sung Kim},
  title = {{SETA}: Lifecycle-Aware Session Execution Tiering in Agentic {LLM} Serving with High-Bandwidth Flash},
  journal = {IEEE Computer Architecture Letters (early access)},
  year = {2026}
}

@inproceedings{petrucci_hbf,
  author = {Vinicius Petrucci and Felippe Zacarias and Vishal Tanna},
  title = {Is High-Bandwidth Flash All You Need?},
  booktitle = {Workshop on Hot Topics in System Infrastructure (HotInfra)},
  year = {2026}
}

@misc{pcie5,
  author = {{PCI-SIG}},
  title = {{PCI Express} Base Specification, Revision 5.0},
  howpublished = {PCI-SIG},
  year = {2019}
}

@misc{micron9550,
  author = {{Micron Technology}},
  title = {Micron 9550 {NVMe} {SSD} Product Brief},
  howpublished = {\url{https://www.micron.com/9550}},
  year = {2024}
}

@misc{gpt5nano,
  author = {{OpenAI}},
  title = {{GPT-5} nano},
  howpublished = {\url{https://platform.openai.com/docs/models/gpt-5-nano}},
  year = {2025}
}

\end{document}